\UseRawInputEncoding

\documentclass[usemulticol,english]{cccconf}
\usepackage[comma,numbers,square,sort&compress]{natbib}
\usepackage{graphicx}          
\usepackage[comma,numbers,square,sort&compress]{natbib}
\usepackage{epstopdf}
\usepackage{mathrsfs}
\usepackage{amsmath}
\usepackage{amssymb}
\usepackage{cuted}
\usepackage[caption=false,font=normalsize,labelfont=sf,textfont=sf]{subfig}
\usepackage{stfloats}
\usepackage{makecell}
\usepackage{color}
\usepackage{multirow}
\usepackage{enumitem}

\usepackage{amsmath,amsfonts}
\usepackage{booktabs}
\usepackage{cellspace}%

\usepackage{makecell}
\setcellgapes{3pt}

\newtheorem{theorem}{Theorem}
\newtheorem{lemma}{Lemma}
\newtheorem{remark}{Remark}
\newtheorem{proposition}{Proposition}
\newtheorem{definition}{Definition}
\newtheorem{assumption}{Assumption}
\newtheorem{learning rule}{Learning rule}
\newtheorem{algorithm}{Algorithm}

\begin{document}

\title{Steady state behavior of the free recall dynamics of  working memory}

\author{Tianhao Li\aref{amss},
        Zhixin Liu*\aref{amss},
        Lizheng Liu\aref{fudan}
        Xiaoming Hu\aref{kth}}

\affiliation[amss]{Key Laboratory of Systems and Control, Academy of Mathematics and Systems Science,Chinese Academy of Sciences, and School of Mathematical Sciences, University of Chinese Academy of Sciences, Beijing 100190, P.~R.~China
        \email{litianhao@amss.ac.cn, lzx@amss.ac.cn}}
\affiliation[fudan]{Academy for Engineering and Technology, Fudan University, Shanghai 200433, P.~R.~China\email{lzliu@fudan.edu.cn}}
\affiliation[kth]{Optimization and Systems Theory, KTH Royal Institute of Technology, 10044 Stockholm, Sweden\email{hu@kth.se}}

\maketitle

\begin{abstract}
This paper studies a dynamical system that models the free recall dynamics of working memory. This model is a modular neural network with $n$ modules, named hypercolumns, and each module consists of $m$ minicolumns. Under mild conditions on the connection weights between minicolumns, we investigate the long-term evolution behavior of the model, namely the existence and stability of equilibriums and limit cycles. We also give a critical value in which Hopf bifurcation happens. Finally, we give a sufficient condition under which this model has a globally asymptotically stable equilibrium with synchronized minicolumn states in each hypercolumn, which implies that in this case recalling is impossible. Numerical simulations are provided to illustrate our theoretical results. A numerical example we give suggests that patterns can be stored in not only equilibriums and limit cycles, but also strange attractors (or chaos).
\end{abstract}

\keywords{Working memory, Free recall, Bifurcation, Asymptotic stability, Strange attractor}

\footnotetext{This work was supported by the National Key R\&D Program of China under Grant 2018YFA0703800, National Natural Science Foundation of
China under Grant 11688101,  Natural Science Foundation of Shandong
Province (ZR2020ZD26), the Strategic Priority Research Program of Chinese Academy of Sciences under Grant No. XDA27000000, and Shanghai Municipal Science and Technology Major Project No. 2021SHZDZX0103.}
\footnotetext{* Corresponding author}

\section{Introduction}
Working memory (WM) is a kind of human memory which is essential for cognitive process and decision making. WM may accept information from sensor memory which represents current environment or long-term memory which represents previous experience. This information is activated persistently and operated mentally in WM \cite{Gazzaniga1998}. As the WM plays a central role in human cognitive process, understanding the mechanisms of WM attracts more and more attention of researchers (e.g. \cite{Brincat2021,Bouchacourt2019,Adam2017}).

Some neural network models are proposed to simulate WM(e.g. \cite{Sandberg2003,Compte2000,Bays2018}), one of which is the attractor neural network model(cf., \cite{Jones2002}). In the attractor neural network, encoded patterns are set as the attractors of the system by regulating connection weights using certain learning rule(cf., \cite{Hopfield1982,Akar2001}). Thus, dynamical network will evolve to the encoded patterns under some conditions. In \cite{Lansner2013}, an attractor neural network WM model constrained by human experimental data on immediate free recall is presented. This model can be written as a compact form of differential equation, which facilitates analyzing dynamics and attractors of the model. The network model consists of $n$ modules where $m$ units are concluded. The module and unit correspond to the hypercolumn and minicolumn in human neural network. Minicolumns in the same hypercolumn interacts via lateral inhibition, such that each hypercolumn acts as a winner-take-all microcircuit\cite{Villani2020}, which means that in a hypercolumn, only one minicolumn is expected to be active. This model can be divided into two parts, learning part and free recall part. In learning part, some patterns are encoded as a sequence of activated minicolumns in the neural network by certain learning rule such as Hebb's rule(cf., \cite{Villani2020,Hebb2002}). In free recall part, the patterns encoded previously are expected to be recalled as the neural network evolves.
The model proposed in \cite{Lansner2013} is employed to simulate the process of a classical experiment, called free recall of word-lists, which is used to explore characteristics of human memory(see, \cite{Kowialiewski2020,Andreasen1995,Howard1999}). In the free recall of word-lists experiment, a list of words are spoken to subjects at a certain rate, and then the subjects are required to write down the words they remember after all the words are spoken(cf.,\cite{Baddeley1974}). ``A large amount of shared variance between immediate free recall and complex WM span task performance has been reported, which implies common mechanisms"\cite{Lansner2013}. This is one of the reasons why the model simulating this experiment is supposed to be able to simulate the WM process.

In \cite{Lansner2013}, simulation results are provided for the behavior of the model introduced above. These results demonstrate that this attractor memory model can remarkably well reproduce key data on immediate free recall of word-lists. Though the learning part of the model in \cite{Lansner2013} is too complicated to analyze mathematically, it is possible to study the free recall part theoretically. In \cite{Villani2020}, G.~Villani et al. study a simplified free recall dynamic of the model where each hypercolumn consists of 2 minicolumns, and gives a sufficient condition for achieving synchronization of positively correlated network units and a necessary condition for having a stable limit cycle as the attractor of each network module. However, G.~Villani et al. require the connection weights of the neural network to have the same absolute value, which is rather restrictive.

In this paper, we focus on the free recall part of the model in \cite{Lansner2013}. Thus we assume that some patterns of memory have already been encoded and consequently the connection weights between minicolumns have been determined. The free recall model we study consists of $n$ hypercolumns where $m$ minicolumns are included. Assumptions on connection weights in this paper are weaker than those in \cite{Villani2020}. In this paper, we firstly show that there is a Hopf bifurcation in the system under some additional assumptions, which indicates the existence of a limit cycle. The latter implies recalling stored patterns. Then, we give a sufficient condition under which this system has a globally asymptotically stable equilibrium consisting of synchronized minicolumn states in each hypercolumn. This case should be avoided in order to recall a pattern.  Finally, we give a numerical example in which the steady state of the free recall model is neither a limit cycle nor an equilibrium, but a strange attractor (or chaos). This example suggests that patterns can also be stored as a strange attractor. Besides being used for determining the steady state behavior and the existence of recalled patterns, our results are also helpful for study of the learning part and learning rules. For example, if the connection weights generated by a learning rule satisfy the condition under which the equilibrium is a globally asymptotically stable equilibrium, then we know that this rule is not an appropriate learning rule since no nontrivial pattern can be recalled.   Difficulty in the theoretical analysis lies in how to analyze the existence of a limit cycle in a $2mn$ dimension state space. Some approaches such as Hopf bifurcation theorem and Lyapunov's methods may be utilized to deal with this difficulty.


The remainder of this paper is organized as follows. We first introduce the free recall model in Section 2. In Section 3, steady state behaviors of the free recall model are discussed. Section 4 gives some simulation results and indicates that the system may converge to a strange attractor (chaos) containing memory patterns.

\section{Problem formulation}

In this section, we will introduce the free recall problem and the model to be studied in this paper.

The model in this paper is the free recall dynamic of the WM model in \cite{Lansner2013}. In the WM model, each memory(or pattern) input is encoded by $n$ attributes (represented by hypercolumns) with $m$ values in each attribute(represented by minicolumns). We can express a pattern by an  $n$ dimension vector, whose $i_{th}$ element value $j$, $j \in \{1,2,\cdots,m\}$, denotes that the $j_{th}$ minicolumn of hypercolumn $i$ is activated. For example, the pattern where the first minicolumn of each hypercolumn is activated can be expressed as $[1,1,...,1]$. Before the free recall process, some patterns are stored in the WM model when the connection weights are determined. These patterns are expected to be recalled in the following free recall process, which means that the output of the free recall model is expected to converge to the corresponding stored patterns.

\begin{figure}[!t]
\centering
\includegraphics[width=2.5in]{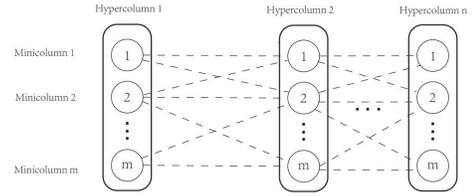}
\caption{The attractor neural network model of free recall.}
\label{fig1}
\end{figure}

Fig.~\ref{fig1} shows an attractor neural network consisting of $n$ hypercolumns where $m$ minicolumns are included. The activity of the minicolumn $j$ in hypercolumn $i$ is represented by its output $o_{ij}$. We say that the minicolumn  $j$ in hypercolumn $i$ is activated if $o_{ij}$ is larger than a threshold that is close to 1. $o_{ij}$ is computed by the following equation (\ref{eq3}) such that a hypercolumn acts as a winner-take-all microcircuit, which means that there is exactly only one minicolumn being activated in each hypercolumn. We use $s_{ij}$ to denote the state of minicolumn $j$ in hypercolumn $i$, and $w_{ij,kl}$ to denote the connection weight from minicolumn $l$ of hypercolumn $k$ to minicolumn $j$ of hypercolumn $i$.

The free recall model can be described by the following dynamics (\cite{Lansner2013,Villani2020}),
\begin{equation}
  \label{eq1}
  \dot{s}_{ij}=\sum_{\substack{k=1 \\ k\neq i}}^{n}\sum_{l=1}^{m}w_{ij,kl}o_{kl}-a_{ij}-s_{ij},
\end{equation}
\begin{equation}
  \label{eq2}
  \tau\dot{a}_{ij}=g_{a}o_{ij}-a_{ij},
\end{equation}
\begin{equation}
  \label{eq3}
  o_{ij}=h(s_{ij})=\frac{e^{s_{ij}}}{\sum_{l=1}^{m}e^{s_{il}}},
\end{equation}
where $s_{ij}\in\mathbb{R}, a_{ij}\in\mathbb{R}$, and $o_{ij}\in(0,1)$ represent the state, level of adaption, and the output of minicolumn $j$ in hypercolumn $i$; $\tau>1$ and $g_{a}>0$ are two constant.

\begin{remark}
   The connection weight $w_{ij,kl}$ in this model is determined by the learning rule and the patterns encoded in the model previously. For example, in the case of $m=2$, which means there are only two minicolumns in each hypercolumn, if we determine connection weights by Hebb's rule(cf., \cite{Villani2020,Hebb2002}), and the patterns we previously store in the model are $[1,1,...,1]$ and $[2,2,...,2]$, then we may get a free recall model where the connection weights satisfy $w_{i1,k1}, w_{i2,k2}>0$, and $w_{i1,k2}, w_{i2,k1}<0$ for $k\neq i$, and $w_{ij,kl}=0$ for $k=i, 1\leq j,l\leq 2$.
\end{remark}

As the attractors of the free recall model are essential to decide whether stored patterns are recalled, it is interesting to discuss what types of attractors occur in the model when the parameters vary.

We first rewrite dynamical model (\ref{eq1})-(\ref{eq3}) in the matrix form,
\begin{equation}
  \label{eq4}
  \left\{
    \begin{array}{c}
    \dot{s}=Wf(s)-s-a,\\
    \dot{a}=\bar{g}_{a}f(s)-\alpha a,
    \end{array}
    \right.
\end{equation}

\begin{equation}
  \label{eq5}
  o=f(s),
\end{equation}

where $\alpha=\frac{1}{\tau}\in (0,1), \bar{g}_{a}=\alpha g_{a}>0$;
\begin{equation}\nonumber
  \begin{aligned}
  &s=[s_{1}^{T},s_{2}^{T},...,s_{n}^{T}]^{T}, s_{i}=[s_{i1},s_{i2},...,s_{im}]^{T}\\
  &a=[a_{1}^{T},a_{2}^{T},...,a_{n}^{T}]^{T}, a_{i}=[a_{i1},a_{i2},...,a_{im}]^{T}\\
  &o=[o_{1}^{T},o_{2}^{T},...,o_{n}^{T}]^{T}, o_{i}=[o_{i1},o_{i2},...,o_{im}]^{T}\\
  &f(s)=[h^{T}(s_{1}),h^{T}(s_{2}),...,h^{T}(s_{n})]^{T},\\
  &h(s_{i})=[h_{1}(s_{i}),h_{2}(s_{i}),...,h_{m}(s_{i})]^{T},\\
  &h_{j}(s_{i})=\frac{e^{s_{ij}}}{\sum_{l=1}^{m}e^{s_{il}}},
  \end{aligned}
\end{equation}
and
$W=(W_{i,k})_{n\times n}$, $W_{i,i}=O_{m\times m}$ and for $k\neq i$,
\begin{equation}
  \nonumber
  W_{i,k}=\begin{bmatrix}
            w_{i1,k1} & \cdots & w_{i1,km} \\
            \vdots & \ddots & \vdots \\
            w_{im,k1} & \cdots & w_{im,km}
          \end{bmatrix}
\end{equation}

For dynamical model (\ref{eq4}), it is difficult to analyze the asymptotic behavior due to the nonlinear and irregular property of the function $f(s)$. In order to simplify the model, we require the following Assumptions 1-3.

\begin{assumption}
  The connection weight matrix $W$ is symmetric.
\end{assumption}
\begin{assumption}
  $\sum_{l=1}^{m}w_{ij,kl}$, the sum of connection weights from all the minicolumns of hypercolumn $k$ to the minicolumns $j$ of hypercolumn $i$, is independent of $j$.
\end{assumption}
We denote
\begin{equation}\label{eq62}
  \lambda_{ik}=\sum_{l=1}^{m}w_{ij,kl},
\end{equation}
and $n\times n$ matrix
\begin{equation}\label{eq63}
  F=(\lambda_{ik})_{n\times n}.
\end{equation}
\begin{assumption}
  $\lambda_{ik}=\lambda_{ki}$, for $1\leq i,k\leq n$.
\end{assumption}

\begin{remark}
  The Assumption 2 is equivalent to $W_{ik}\vec{\mathbf{1}}_{m}=\lambda_{ik}\vec{\mathbf{1}}_{m}$, where $\vec{\mathbf{1}}_{m}=[1,1,\cdots,1]^{T}$ is an $m$ dimension vector.
\end{remark}

In the next section, we will analyze dynamical behaviors of the system (\ref{eq4}).

\section{Analysis of the dynamical system}
We first consider equilibria of the model (\ref{eq4}), which satisfy the following equation,
\begin{equation}\label{eq6}
 (W-\frac{\bar{g}_{a}}{\alpha}I)f(s)=s, a=\frac{\bar{g}_{a}}{\alpha}f(s),
\end{equation}
where $W$ is an $mn\times mn$ matrix, $s$ and $a$ are $mn$ vector.

It is clear that $s_{0}=[c_{1}\vec{\mathbf{1}}^{T}_{m},c_{2}\vec{\mathbf{1}}^{T}_{m},\cdots,c_{n}\vec{\mathbf{1}}^{T}_{m}]^{T}, a_{0}=\frac{\bar{g}_{a}}{\alpha}[\vec{\mathbf{1}}^{T}_{m},\vec{\mathbf{1}}^{T}_{m},\cdots,\vec{\mathbf{1}}^{T}_{m}]^{T}$ is a solution of (\ref{eq6}), where $c_{i}=(\sum_{k=1}^{n}\lambda_{ik}-\bar{g}_{a}/\alpha)/m, \vec{\mathbf{1}}^{T}_{m}=[1,1,\cdots,1]^{T}\in \mathbb{R}^{m}$.

Moving the equilibrium $[s_{0}^{T},a_{0}^{T}]^{T}$ of the model (\ref{eq4}) to the origin by the transformation $\bar{s}=s-s_{0},\bar{a}=a-a_{0}$, by (\ref{eq5}), we obtain the following equation,
\begin{equation}
    \label{eq7}
    \begin{aligned}
    &\left\{
        \begin{array}{c}
        \dot{\bar{s}}=W\bar{f}(\bar{s})-\bar{s}-\bar{a},\\
        \dot{\bar{a}}=\bar{g}_{a}\bar{f}(\bar{s})-\alpha \bar{a},
        \end{array}
        \right. \\
    &o=f(\bar{s}),
    \end{aligned}
\end{equation}
where $\bar{f}(\bar{s})=f(\bar{s})-\frac{1}{m}\vec{\mathbf{1}}^{T}_{mn}$.

For the function $\bar{f}(\bar{s})$, we have the following lemma.

\begin{lemma}
  $\bar{s}^{T}\bar{f}(\bar{s})\geq 2\bar{f}^{T}(\bar{s})\bar{f}(\bar{s})\geq 0, \forall \bar{s} \in \mathbb{R}^{mn}.$
\end{lemma}
Proof: By calculation, we have
\begin{equation}\nonumber
  \begin{aligned}\nonumber
    &\sum_{j=1}^{m}\sum_{r=1}^{m}\bar{s}_{ij}(e^{\bar{s}_{ij}}-e^{\bar{s}_{ir}})\\
    &=\sum_{1\leq j<r\leq m}\bar{s}_{ij}(e^{\bar{s}_{ij}}-e^{\bar{s}_{ir}})+\sum_{1\leq r<j\leq m}\bar{s}_{ij}(e^{\bar{s}_{ij}}-e^{\bar{s}_{ir}})\\
    &=\sum_{1\leq j<r\leq m}\bar{s}_{ij}(e^{\bar{s}_{ij}}-e^{\bar{s}_{ir}})+\sum_{1\leq j<r\leq m}\bar{s}_{ir}(e^{\bar{s}_{ir}}-e^{\bar{s}_{ij}})\\
    &=\sum_{1\leq j<r\leq m}(\bar{s}_{ij}-\bar{s}_{ir})(e^{\bar{s}_{ij}}-e^{\bar{s}_{ir}}).
\end{aligned}
\end{equation}
Similarly,
\begin{equation}\nonumber
  \sum_{j=1}^{m}\sum_{r=1}^{m}(e^{\bar{s}_{ij}}-e^{\bar{s}_{it}})(e^{\bar{s}_{ij}}-e^{\bar{s}_{ir}})=\sum_{1\leq j<r\leq m}(e^{\bar{s}_{ij}}-e^{\bar{s}_{ir}})^{2}.
\end{equation}
Thus, we have
\begin{equation}\label{eq24}
\begin{aligned}
    &\bar{s}^{T}\bar{f}(\bar{s})\\
    &=\sum_{i=1}^{n}\frac{\sum_{j=1}^{m}\sum_{r=1}^{m}\bar{s}_{ij}(e^{\bar{s}_{ij}}-e^{\bar{s}_{ir}})}{m\sum_{l=1}^{m}e^{\bar{s}_{il}}}\\
    &=\sum_{i=1}^{n}\frac{\sum_{1\leq j<r\leq m}(\bar{s}_{ij}-\bar{s}_{ir})(e^{\bar{s}_{ij}}-e^{\bar{s}_{ir}})}{m\sum_{l=1}^{m}e^{\bar{s}_{il}}},
\end{aligned}
\end{equation}
and
\begin{equation}\label{eq25}
\begin{aligned}
    &\bar{f}^{T}(\bar{s})\bar{f}(\bar{s})\\
    &=\sum_{i=1}^{n}\frac{\sum_{t=1}^{m}\sum_{j=1}^{m}\sum_{r=1}^{m}(e^{\bar{s}_{ij}}-e^{\bar{s}_{it}})(e^{\bar{s}_{ij}}-e^{\bar{s}_{ir}})}{(m\sum_{l=1}^{m}e^{\bar{s}_{il}})^{2}}\\
    &=\sum_{i=1}^{n}\frac{\sum_{1\leq j<r\leq m}(e^{\bar{s}_{ij}}-e^{\bar{s}_{ir}})^{2}}{m(\sum_{l=1}^{m}e^{\bar{s}_{il}})^{2}}\\
    &\geq 0.
\end{aligned}
\end{equation}
Besides, we have
\begin{equation}\label{eq26}
\begin{aligned}
   \frac{\mid e^{\bar{s}_{ij}}-e^{\bar{s}_{ir}}\mid}{\sum_{k=1}^{m}e^{\bar{s}_{ik}}}
  & \leq \frac{\mid e^{\bar{s}_{ij}}-e^{\bar{s}_{ir}}\mid}{e^{\bar{s}_{ij}}+e^{\bar{s}_{ir}}} \\
  & =\mid tanh(\frac{\bar{s}_{ij}-\bar{s}_{ir}}{2})\mid \\
  & \leq \frac{\mid \bar{s}_{ij}-\bar{s}_{ir}\mid}{2}.
\end{aligned}
\end{equation}

By the equations (\ref{eq24})-(\ref{eq26}), we obtain the conclusion of this lemma. $\blacksquare$

Denote the largest eigenvalue of the connection weight matrix $W$ by $\mu_{max}(W)$. It is clear that under the condition of $\bar{g}_{a}>\alpha (\mu_{max}(W)-2)$, $\bar{s}=0$ is the unique equilibrium of the system (\ref{eq7}). If the system (\ref{eq7}) has another equilibrium $\bar{s}=\bar{s}_{0}$ different from the origin, then
\begin{equation}
  \label{13}
  (W-\frac{\bar{g}_a}{\alpha}I)\bar{f}(\bar{s}_{0})=\bar{s}_{0}.
\end{equation}
By (\ref{13}), we have
\begin{equation}
  \label{14}
  \bar{s}_{0}^{T}\bar{f}(\bar{s}_{0})\leq(\mu_{max}(W)-\frac{\bar{g}_a}{\alpha})\bar{f}^{T}(\bar{s}_{0})\bar{f}(\bar{s}_{0}).
\end{equation}
By (\ref{13}) and the assumption $\bar{s}_{0}\neq 0$, we have $\bar{f}(\bar{s}_{0})\neq 0$.

If the condition $\bar{g}_{a}>\alpha (\mu_{max}(W)-2)$ is satisfied, then we get a contradiction between (\ref{14}) and Lemma 1. By reduction to absurdity, we have the following proposition.

\begin{proposition}
If $\bar{g}_{a}>\alpha (\mu_{max}(W)-2)$, then $\bar{s}=0$ is the unique equilibrium of the system (\ref{eq7}).
\end{proposition}

\subsection{Bifurcation analysis}

``The Hopf bifurcation refers to the development of
periodic orbits (`self-oscillations') from a stable fixed
point, as a parameter crosses a critical value"\cite{Marsden1976}. We will use Hopf bifurcation theorem to prove the existence of a limit cycle attractor in the system (\ref{eq7}) under some conditions. ``Vague attractor" (cf., \cite{Marsden1976}) is introduced  to prove the asymptotic stability of the limit cycle.

We know that for the system where the Hopf bifurcation occurs, a pair of complex conjugate eigenvalues of Jacobian matrix at an equilibrium crosses imaginary axis of complex plane when the bifurcation parameter crosses the bifurcation value. Consider the linearized system of (\ref{eq7}) at the origin,
\begin{equation}
  \label{eq8}
  \dot{y}=Hy,
\end{equation}
where $y=[\bar{s}^{T},\bar{a}^{T}]^T$, the corresponding Jacobian matrix at the origin is
\begin{equation}
  \label{eq9}
  H =\left[
  \begin{array}{cc}
  W\Lambda-I & -I \\
  \bar{g}_{a}\Lambda & -\alpha I
  \end{array}
  \right],
\end{equation}
\begin{equation}
  \label{eq10}
  \Lambda =\frac{\partial \bar{f}}{\partial \bar{s}}|_{\bar{s}=0}=
  \begin{bmatrix}
    \bar{\Lambda} &  &  \\
     & \ddots &  \\
     &  & \bar{\Lambda}
  \end{bmatrix},
\end{equation}
$\bar{\Lambda}=\frac{1}{m}I-\frac{1}{m^{2}}X$ and $X=\vec{\mathbf{1}}_{m}\vec{\mathbf{1}}^{T}_{m}$.

To calculate eigenvalues of the matrix $H$, we need the following lemma.

\begin{lemma}
  Under Assumptions 1-3, there exists an $mn \times mn$ orthogonal matrix $P=[p_{1},p_{2},\cdots, p_{mn}]$, such that

   (\romannumeral1) $span\{p_{(m-1)n+k}|1\leq k\leq n\}=span\{\vec{\mathbf{\varepsilon}}_{k}\otimes \vec{\mathbf{1}}_{m}|1\leq k \leq n\}$,

   (\romannumeral2)\begin{equation}\nonumber
           P^{T}W\Lambda P=\frac{1}{m}
                        \begin{bmatrix}
                        \mu_{1} &  &  &  &  &  \\
                         & \ddots &  &  &  &  \\
                         &  & \mu_{(m-1)n} &  &  &  \\
                         &  &  & 0 &  &  \\
                         &  &  &  & \ddots &  \\
                         &  &  &  &  & 0
                      \end{bmatrix}\triangleq D_{1},
      \end{equation}

   (\romannumeral3)\begin{equation}\nonumber
           P^{T}\Lambda P=\frac{1}{m}
                     \begin{bmatrix}
                       I_{(m-1)n} &  \\
                        & O_{n}
                     \end{bmatrix}\triangleq D_{2},
      \end{equation}

   (\romannumeral4)\begin{equation}\nonumber
      \begin{aligned}
        P^{T}WP &= \begin{bmatrix}
                        \mu_{1} &  &  &  &  &  \\
                         & \ddots &  &  &  &  \\
                         &   & \mu_{(m-1)n} &  &   &   \\
                          &   &   & \mu_{(m-1)n+1} &   &   \\
                          &   &   &   & \ddots &   \\
                          &   &   &   &   & \mu_{mn}
                      \end{bmatrix} \\
         & \triangleq D_{3},
      \end{aligned}
      \end{equation}
   where $\otimes$ is the kronecker product, $\vec{\mathbf{\varepsilon}}_{k}$ is $k_{th}$ column of $n$ dimension unit matrix, $\{\frac{\mu_{r}}{m}|1\leq r\leq (m-1)n\}$ are eigenvalues of $W \Lambda$ in descending order besides 0 eigenvalue of $n$ multiplicity, $\{\mu_{r}|  (m-1)n+1\leq r\leq mn\}$ are eigenvalues of $W$ in descending order besides $\{\mu_{r}|1\leq r\leq (m-1)n\}$.
\end{lemma}

Proof: By Assumptions 1-3, it is clear that $W\Lambda$ is real and symmetric, and has $0$ eigenvalue of $n$ multiplicity with eigenvectors of $\vec{\mathbf{\varepsilon}}_{k}\otimes \vec{\mathbf{1}}_{m} (1\leq k \leq n)$. Thus, we can define an $mn \times mn$ orthogonal matrix $\tilde{P}=[\tilde{p}_{1},\tilde{p}_{2},\cdots, \tilde{p}_{mn}]$, where $\tilde{p}_{(m-1)n+k}=\vec{\mathbf{\varepsilon}}_{k}\otimes \vec{\mathbf{1}}_{m}$, and $\tilde{p}_{r} (1\leq r \leq (m-1)n)$ is the eigenvector of $W\Lambda$ corresponding to eigenvalue $\frac{\mu_{r}}{m}$. It is clear that conclusions (\romannumeral1) (\romannumeral2) and (\romannumeral3) are correct if matrix $P$ is replaced by $\tilde{P}$. Furthermore, by the orthogonality of $\tilde{P}$ and Remark 2, we have
\begin{equation}\nonumber
    W\tilde{p}_{r}=m W\Lambda \tilde{p}_{r}=\mu_{r}\tilde{p}_{r}, \qquad 1\leq r\leq (m-1)n,
\end{equation}
\begin{equation}\nonumber
    W\tilde{p}_{(m-1)n+k} = \sum_{i=1}^{n}\lambda_{ik}\tilde{p}_{(m-1)n+i}, \qquad 1\leq k\leq n,
\end{equation}
where $\lambda_{ik}$ is defined in (\ref{eq62}).
Thus, we have
\begin{equation}\label{eq95}
  \tilde{P}^{T}W \tilde{P}=\begin{bmatrix}
           \mu_{1} &   &   &   \\
             & \ddots &   &   \\
             &   & \mu_{(m-1)n} &   \\
             &   &   & F
         \end{bmatrix},
\end{equation}
where $n\times n$ matrix $F$ is defined in (\ref{eq63}). Define $P=\tilde{P}\hat{P}$, where
\begin{equation}\nonumber
  \hat{P}=\begin{bmatrix}
            I_{(m-1)n} &   \\
              & \hat{P}_{0}
          \end{bmatrix},
\end{equation}
and $\hat{P}_{0}$ is a $n\times n$ orthogonal matrix which satisfies
\begin{equation}\nonumber
  \hat{P}_{0}^{T}F\hat{P}_{0}=\begin{bmatrix}
                                \mu_{(m-1)n+1} &   &   &   \\
                                  & \mu_{(m-1)n+2} &   &   \\
                                  &   & \ddots &   \\
                                  &   &   & \mu_{mn}
                              \end{bmatrix}.
\end{equation}
Then $P$ is the $mn\times mn$ orthogonal matrix we need. $\blacksquare$

\begin{remark}
 Denote
 \begin{equation}\nonumber
   p_{r}=[p_{r,11},p_{r,12},\cdots,p_{r,1m},p_{r,21},\cdots,p_{r,nm}]^{T}.
 \end{equation}
 By the fact that $P$ is orthogonal and conclusion (\romannumeral1) of Lemma 2, we have $\sum_{l=1}^{m}p_{r,kl}=0$, for $1\leq r\leq (m-1)n, 1\leq k\leq n$. Furthermore, for $(m-1)n+1\leq r\leq mn, 1\leq k\leq n$, we obtain that $p_{r,kl}$ is independent of $l$. Thus, we can denote $p_{r,kl}$ by $c_{r,k}$ for $(m-1)n+1\leq r\leq mn, 1\leq k\leq n$.
\end{remark}

Next, we calculate eigenvalues of $H$.
By Lemma 2, we have
\begin{equation}
  \label{eq11}
  H = \left[\begin{array}{cc}P & O \\ O & P\end{array}\right] \left[\begin{array}{cc} D_{1}-I & -I \\ \bar{g}_{a}D_{2} & -\alpha I\end{array}\right] \left[\begin{array}{cc} P^{T} & O \\ O & P^{T}\end{array}\right].
\end{equation}

Denote the eigenvalues of $H$ as $\nu_{i,\pm} (1\leq i \leq mn)$. From (\ref{eq11}), we have for $1\leq i\leq (m-1)n$
\begin{equation}
  \label{eq90}
  \nu_{i,\pm} = \frac{\mu_{i}-m(1+\alpha)}{2m} \pm \frac{1}{2}\sqrt{(\alpha+(\frac{\mu_{i}}{m}-1))^2-\frac{4\bar{g}_a}{m}},
\end{equation}
and for $(m-1)n+1\leq i\leq mn$
\begin{equation}
  \label{eq12}
  \nu_{i,+} = -\alpha, \nu_{i,-} = -1.
\end{equation}

Based on the above analysis, we have the following lemma.

\begin{lemma}

(1) If $\mu_{1}\leq m(1-\alpha)$, then
   \begin{equation}\nonumber
     \max \limits_{1\leq i\leq (m-1)n}Re(\nu_{i,\pm})\leq -\alpha.
   \end{equation}
(2) If
\begin{equation}
  \nonumber
\mu_{1}> m(1-\alpha), \bar{g}_{a}>\frac{m(\frac{\mu_{1}}{m}+\alpha-1)^{2}}{4},
\end{equation}
then
\begin{equation}
  \nonumber
  \max \limits_{1\leq i\leq (m-1)n}Re(\nu_{i,\pm})=\frac{\mu_{1}-m(1+\alpha)}{2m}.
\end{equation}
\end{lemma}

By Lemma 3, we know that if $\mu_{1}=m(1+\alpha)$, and $\bar{g}_{a}>m\alpha^2$, then
\begin{equation}
  \nonumber
  \max \limits_{1\leq i\leq (m-1)n}Re(\nu_{i,\pm})=0.
\end{equation}

In order to analyze the dynamical behavior of the system (\ref{eq7}), we need the following theorem in \cite{Marsden1976}.

For the system $\dot{v}=X_{\mu}(v)$, where $v \in \mathbb{R}^{n}$ is the state, $\mu \in \mathbb{R}$ is a parameter, and $X_{\mu}$ is a $k+1(k\geq 4)$ times continuously differentiable ($C^{k+1}$) vector field on $\mathbb{R}^{n}$. Let $X_{\mu}$ satisfy that $X_{\mu}(0)=0$ for all $\mu$.
\begin{theorem}(\cite{Marsden1976})If

  (\romannumeral1) $D_{v}X_{\mu}(0)$ has two distinct, complex conjugate eigenvalues $\lambda(\mu)$ and $\bar{\lambda}(\mu)$ such that $Re(\lambda(0))=0$ and for $\mu > 0$ $Re(\lambda(\mu))>0$, and the rest of the spectrum are distinct from $\lambda(\mu)$ and $\bar{\lambda}(\mu)$,

  (\romannumeral2)
  \begin{equation}\nonumber
   \left. \frac{dRe(\lambda(\mu))}{d\mu}\right|_{\mu=0}>0,
  \end{equation}

\noindent  then there exist two neighborhoods of 0 in $\mathbb{R}$, denoted by $(-\varepsilon,\varepsilon)$ and $\bar{N}$, a unique $C^{k-2}$ function $\mu : (-\varepsilon,\varepsilon)\rightarrow \bar{N}$ and a continues function $v_{3} : (-\varepsilon,\varepsilon)\rightarrow \mathbb{R}^{n-2}$ such that for any parameter $\mu$ satisfying $\mu = \mu(v_{1}), v_{1} \in (-\varepsilon,\varepsilon)$, the system has a closed orbit. Furthermore, point $(v_{1},0,v_{3}(v_{1}))$ is on this closed orbit.
\end{theorem}

Theorem 1 gives a sufficient condition under which a closed orbit bifurcate from the origin in a system with a parameter. In \cite{Marsden1976}, the definition of ``vague attractor" is introduced to discuss the uniqueness and asymptotic stability of the closed orbit above.

Let $X_{\mu}=[X^{1}_{\mu}, X^{2}_{\mu}, (X^{3}_{\mu})^{T}]^{T}$, where $X^{1}_{\mu}$ and $X^{2}_{\mu}$ are coordinates in the eigenspace of $D_{v}X_{0}(0)$ corresponding to the eigenvalues $\lambda(0)$ and $\bar{\lambda}(0)$, and $X^{3}_{\mu}$ is a coordinate in a subspace complementary to this eigenspace. Correspondingly, we denote $v = [v_{1},v_{2},v_{3}^{T}]^{T}$. Under the condition of Theorem 1, the coordinate system can be chosen to satisfy (cf.,\cite{Marsden1976})
  \begin{equation}\label{eq13}
    D_{v}X_{0}(0)=\left[\begin{array}{ccc}
                      0 & \mid\lambda(0)\mid & 0 \\
                      -\mid\lambda(0)\mid & 0 & 0 \\
                      0 & 0 & \frac{\partial X_{0}^{3}}{\partial v_{3}}(0)
                    \end{array}\right].
  \end{equation}

By the center manifold theorem there is a center manifold for the flow of the system $\left[\begin{array}{c}\dot{v} \\ \dot{\mu}\end{array}\right] = \left[\begin{array}{c}X_{\mu} \\ 0\end{array}\right]$ tangent to the eigenspace of $\lambda(0)$ and $\bar{\lambda}(0)$, and to the $\mu$-axis at the point $\left[\begin{array}{c}v \\ \mu\end{array}\right] = \left[\begin{array}{c}0 \\ 0\end{array}\right]$. The center manifold may be represented locally as the graph of a function $g$, that is, as $[v_{1},v_{2},g^{T}(v_{1},v_{2},\mu),\mu]^{T}$ for $[v_{1},v_{2},\mu]^{T}$ in some neighborhood of $[0,0,0]^{T}$. Also, $g(0,0,0)=dg(0,0,0)=0$. Define $\hat{X}_{\mu}(v_{1},v_{2})=(X^{1}_{\mu}(v_{1},v_{2},g^{T}(v_{1},v_{2},\mu)),X^{2}_{\mu}(v_{1},v_{2},g^{T}(v_{1},v_{2},\mu)))$, then $\hat{X}_{\mu}$ is a smooth one-parameter family of vector fields in some neighborhood of the origin on the 2-dimension center manifold because the center manifold is locally invariant under the flow of $X=\left[\begin{array}{c}X_{\mu} \\ 0\end{array}\right]$. A function $\tilde{V}(v_{1})$ is defined in some neighborhood of the origin in $\mathbb{R}$ by the following equation (\ref{eq14}) for the flow of $\hat{X}_{0}$ when the bifurcation parameter $\mu=0$,
  \begin{equation}\label{eq14}
    \tilde{V}(v_{1})=\int_{0}^{T(v_{1})}\hat{X}_{0}^{1}(b_{t}(v_{1},0),c_{t}(v_{1},0))dt,
  \end{equation}
where $[b_{t},c_{t}]^{T}$ is the flow of $\hat{X}_{0}$, $T(v_{1})$ is the minimum time $t>0$ when $c_{t}(v_{1},0)=0$ and $b_{t}(v_{1},0)$ has the same sign with $v_{1}$.

\begin{definition}
  $v=0$ is called a ``vague attractor" for the flow of $X_{0}(v)$, if $\frac{d^{3}\tilde{V}}{dv_{1}^{3}}(0)<0$, where $\tilde{V}$ is defined in the equation (\ref{eq14}).
\end{definition}

\begin{theorem}(\cite{Marsden1976})
Let the condition of Theorem 1 be satisfied. If

  (\romannumeral1) the rest of spectrum of $D_{v}X_{\mu}(0)$ besides $\lambda(\mu)$ and $\bar{\lambda}(\mu)$ remain in the left half complex plain as $\mu$ crosses zero,

\noindent  then we have

  (A) there is a neighborhood $U$ of $(v,\mu)=(0,0)$ in $\mathbb{R}^{n+1}$ such that any closed orbit in $U$ is one of those in the conclusion of Theorem 1.

\noindent Furthermore, if

  (\romannumeral2) $v=0$ is a ``vague attractor" for the flow of $X_{0}(v)$,

\noindent then we have

  (B) $\mu(v_{1})>0$ for all $v_{1} \in (-\varepsilon, 0)\bigcup (0,\varepsilon)$ and the closed orbit is asymptotically stable.
\end{theorem}

We next give a theorem based on Theorem 1 and Theorem 2.

\begin{theorem}
   For the system (\ref{eq7}), we suppose that Assumptions 1-3 are satisfied. $\mu_{1}$ is the largest eigenvalue of $W\Lambda$ besides 0, $W$ is the connection weight matrix and $\Lambda$ is defined in (\ref{eq10}). If
  \begin{itemize}
    \item [(1)] $\mu = \mu_{1}$ is a simple eigenvalue of $W\Lambda$, and $$\bar{g}_{a}>\frac{m(\frac{\mu_{1}}{m}+\alpha-1)^{2}}{4},$$
    \item [(2)] one of the following conditions is satisfied,
             \begin{itemize}[leftmargin=20pt,topsep=5pt]
               \item [(a)]  $m=2$;
               \item [(b)]  $m=3$ and $\bar{g}_{a}\geq m(1+\alpha)^{2}$;
                  \item [(c)]  $m\geq 4$, $\bar{g}_{a}\geq m(1+\alpha)^{2}$, and
             \begin{equation}\nonumber
                 \frac{3}{m}\sum_{r=1}^{n}(\sum_{t=1}^{m}p^{2}_{1,rt})^{2} \geq \sum_{r=1}^{n}\sum_{t=1}^{m}p^{4}_{1,rt},
             \end{equation}
             \end{itemize}
  \end{itemize}
  where $p_{1}=[p_{1,11},p_{1,12},\cdots,p_{1,1m},p_{1,21},p_{1,22},\cdots,p_{1,mn}]^{T}$ is the eigenvector of $W\Lambda$ corresponding to the eigenvalue $\mu_{1}$,
  then there exists a real number $\delta >0$ such that for any $\mu_{1} \in (m(1+\alpha),m(1+\alpha)+\delta)$, there is a neighborhood of the origin in $\mathbb{R}^{2mn}$ which contains a closed orbit for the system (\ref{eq7}). Furthermore, this closed orbit is unique in the above neighborhood and is orbitally asymptotically stable.

\end{theorem}

Proof: Theorem 1 and Conclusion (B) of Theorem 2 are used to prove the existence of the closed orbit. Theorem 2 is used to prove the uniqueness and asymptotic stability of the closed orbit. Conclusion (A) of Theorem 2 guarantees the uniqueness of the closed orbit. Conclusion (B) of Theorem 2 guarantees that the closed orbit is asymptotically stable. Thus, we need to show that the conditions of Theorem 1 and Theorem 2 can be satisfied. Define $\kappa=\mu_{1}-m(1+\alpha)$ as the bifurcation parameter, and denote the two distinct complex conjugate eigenvalues of the matrix $H$ defined in the equation (\ref{eq9}) with the largest real part as $\lambda(\kappa)$ and $\bar{\lambda}(\kappa)$. We have $Re(\lambda(\kappa))=\frac{\kappa}{2m}$ by Lemma 3.
At the critical value $\kappa =0$, the following two conditions are satisfied.

\begin{itemize}
\item [(1)] $Re(\lambda(0))=0$, and $Re(\lambda(\kappa))>0$ if $\kappa>0$. By (\ref{eq90}) and (\ref{eq12}), the rest of the spectrum of $H$ are distinct from $\lambda(\kappa)$ and $\bar{\lambda}(\kappa)$ as $\mu = \mu_{1}$ is a simple eigenvalue of $W\Lambda$.

\item [(2)]
\begin{equation}
  \nonumber
  \left.\frac{dRe(\lambda(\kappa))}{d\kappa}\right|_{\kappa=0}=\frac{1}{2m}>0.
\end{equation}
\end{itemize}
Thus, the conditions of Theorem 1 are verified. We next verify the conditions of Theorem 2.
\begin{itemize}
\item [(3)] The condition (\romannumeral1) of Theorem 2 is satisfied as the real part of $\lambda(\kappa)$ and $\bar{\lambda}(\kappa)$ is strictly larger than those of the other eigenvalues of matrix $H$ when $\mu = \mu_{1}$ is a simple eigenvalue of $W\Lambda$.

\item [(4)] To show that the condition (\romannumeral2) of Theorem 2 is satisfied, we need to prove $\frac{d^{3}\tilde{V}}{dv_{1}^{3}}(0)<0$ when one of  conditions (a) (b) and (c) is satisfied, where $\tilde{V}$ is defined in (\ref{eq14}). The proof is given in Appendix B.
\end{itemize}

This completes the proof.                 $\blacksquare$

\subsection{Equilibrium analysis}

We will analyze the stability of the origin of the system (\ref{eq7}). By Lemma 3 and Lyapunov's first method, we know that if $\bar{g}_{a}>\frac{m(\frac{\mu_{1}}{m}+\alpha-1)^{2}}{4}$, then the origin is an unstable equilibrium when $\mu_{1}>m(1+\alpha)$ and locally exponentially stable equilibrium when $\mu_{1}<m(1+\alpha)$. Moreover, under some conditions, we have globally asymptotic stability of the origin equilibrium of the system (\ref{eq7}).

We first introduce the following lemmas.

\begin{lemma}(\cite{Spivak2018})
  If $\frac{\partial \bar{f}}{\partial \bar{s}}$ is symmetric for any $\bar{s}\in \mathbb{R}^{mn}$, then there exists a function $\bar{F}: \mathbb{R}^{mn}\rightarrow \mathbb{R}$ such that $\frac{\partial \bar{F}}{\partial \bar{s}}=\bar{f}(\bar{s})$, and the value of the integral $\int_{0}^{\bar{s}}\bar{f}(u)du$ is path independent.
\end{lemma}

\begin{lemma}
   $\int_{0}^{\bar{s}}\bar{f}(u)du\geq 0, \forall \bar{s}\in \mathbb{R}^{mn}.$
\end{lemma}
Proof: By Lemma 4 and the fact that $\frac{\partial \bar{f}}{\partial \bar{s}}$ is symmetric, the value of the integral $\int_{0}^{\bar{s}}\bar{f}(u)du$ is path independent. Thus, this integral is well-defined. We calculate the integral along the straight line between the origin and $\bar{s}$. By Lemma 1, we have
\begin{equation}\nonumber
  \int_{0}^{\bar{s}}\bar{f}(u)du = \int_{0}^{1}\bar{s}^{T}\bar{f}(t\bar{s})dt \geq 0.
\end{equation}
$\blacksquare$

By the above two lemmas, we have the following proposition.

\begin{table*}[!htb]
\centering\makegapedcells
  \caption{Dynamics of the system (\ref{eq7})}
  \label{tab1}
  \scalebox{0.93}{
  \begin{tabular}{c|c|c|c}
    \hline
      &globally asymptotically stable equilibrium& asymptotically stable equilibrium & asymptotically stable limit cycle \\ \hline 
    $m=2$&\multirow{3}{*}[-40pt]{\makecell*[c]{$\displaystyle \mu_{max}(W)<2(1+\alpha),$
    \\[2ex] $\displaystyle \bar{g}_{a}>\dfrac{2\alpha^{2} (1+\alpha)\|W\|^2}{(2(1+\alpha)-\mu_{max}(W))^2}$}} & \multirow{3}{*}[-40pt]{\makecell*[c]{$\displaystyle \mu_{1}<m(1+\alpha)$,\\[2ex]$\displaystyle \bar{g}_{a}>m\alpha^{2}$}} & \makecell*[c]{$\displaystyle m(1+\alpha)<\mu_{1}<m(1+\alpha)+\delta$,\\[1ex]  
    $\displaystyle \bar{g}_{a}>\dfrac{m(\dfrac{\mu_{1}}{m}+\alpha-1)^{2}}{4}$ } \\ \cline{1-1} \cline{4-4}
    $m=3$& &  & \makecell*[c]{$\displaystyle m(1+\alpha)<\mu_{1}<m(1+\alpha)+\delta$,\\[1ex] $\displaystyle \bar{g}_{a}>\max\{\dfrac{m(\dfrac{\mu_{1}}{m}+\alpha-1)^{2}}{4},
    m(1+\alpha)^{2}\}$}
    \\ \cline{1-1} \cline{4-4} 
    $m\geq 4$ & & & \makecell*[c]{$\displaystyle m(1+\alpha)<\mu_{1}<m(1+\alpha)+\delta$,\\[1ex] $\displaystyle \bar{g}_{a}>\max\{\dfrac{m(\dfrac{\mu_{1}}{m}+\alpha-1)^{2}}{4},
    m(1+\alpha)^{2}\}$,\\[1ex]
    $\displaystyle \dfrac{3}{m}\sum_{r=1}^{n}(\sum_{t=1}^{m}p^{2}_{1,rt})^{2} \geq \sum_{r=1}^{n}\sum_{t=1}^{m}p^{4}_{1,rt}$} \\[1ex]
    \hline
  \end{tabular}
  }
\end{table*}

\begin{proposition}
If $\mu_{max}(W)<2(1+\alpha)$, and $\bar{g}_{a}$ satisfies the following equation,
\begin{equation}
  \label{eq30}
  \bar{g}_{a}>\frac{2\alpha^{2} (1+\alpha)\|W\|^2}{\sigma^2},
\end{equation}
where $\sigma = 2(1+\alpha)-\mu_{max}(W)$, then the origin is a globally asymptotically stable equilibrium of the system (\ref{eq7}).
\end{proposition}

Proof: We introduce the following Lyapunov function to prove this proposition,
\begin{equation}
\begin{aligned}\nonumber
  V(\bar{s},\bar{a})&=g_{a}\int_{0}^{\bar{s}}\bar{f}(u)du +\frac{(1-\varepsilon)(1+\alpha)}{2}\bar{s}^{T}\bar{s}\\
  &-(1-\varepsilon)\bar{s}^{T}\bar{a}+\frac{1}{2\alpha}\bar{a}^{T}\bar{a}.
\end{aligned}
\end{equation}
where $g_{a}=\bar{g}_{a}/\alpha$, and $\varepsilon \in (0,1)$ is a parameter to be determined.

By Lemma 4 and the fact that $\frac{\partial \bar{f}}{\partial \bar{s}}$ is symmetric, the function $V(\bar{s},\bar{a})$ is well-defined.

By Lemma 5, we know that $V(\bar{s},\bar{a})$ is a positive definite and radially unbounded function.

The derivative of $V(\bar{s},\bar{a})$ can be calculated as follows,
\begin{equation}
\begin{aligned}\label{eq31}
  \dot{V}&=g_{a}(\bar{f}^{T}(\bar{s})W\bar{f}(\bar{s})-(1+(1-\varepsilon)\alpha)\bar{s}^{T}\bar{f}(\bar{s}))\\
    &+(1-\varepsilon)(1+\alpha)\bar{s}^{T}W\bar{f}(\bar{s})-(1-\varepsilon)(1+\alpha)\bar{s}^{T}\bar{s}\\
    &-(1-\varepsilon)\bar{a}^{T}W\bar{f}(\bar{s})-\varepsilon \bar{a}^{T}\bar{a}.
\end{aligned}
\end{equation}
By the equation (\ref{eq31}), we have
\begin{equation}
\begin{aligned}\label{eq32}
\dot{V}&\leq g_{a}(\mu_{max}(W)\bar{f}^{T}(\bar{s})\bar{f}(\bar{s})-(1+(1-\varepsilon)\alpha)\bar{s}^{T}\bar{f}(\bar{s}))\\
    &+(1-\varepsilon)(1+\alpha)\|W\|\|\bar{s}\|\|\bar{f}(\bar{s})\|-(1-\varepsilon)(1+\alpha)\bar{s}^{T}\bar{s}\\
    &+(1-\varepsilon)\|W\|\|\bar{a}\|\|\bar{f}(\bar{s})\|-\varepsilon \bar{a}^{T}\bar{a},
\end{aligned}
\end{equation}
where $\|\cdot\|$ denotes the Euclidean norm of matrixes and vectors.

We introduce two parameters $\gamma$ and $\delta$ which satisfy the following equations (\ref{eq33}) and (\ref{eq34})
\begin{equation}
  \label{eq33}
  \varepsilon\alpha+\gamma+\delta = \frac{\sigma}{2}, \gamma >0, \delta>0,
\end{equation}
\begin{equation}\label{eq34}
  \delta = \frac{\sigma}{4}, \gamma =\varepsilon \delta.
\end{equation}
By the equations (\ref{eq33}) and (\ref{eq34}), we have
\begin{equation}\label{eq35}
  \varepsilon = \frac{\sigma}{\sigma+4\alpha}, \gamma = \frac{\sigma^{2}}{4(\sigma+4\alpha)}.
\end{equation}

By the equation (\ref{eq32}), we have
\begin{equation}
\begin{aligned}\label{eq36}
\dot{V}&\leq g_{a}\left(\mu_{max}(W)\bar{f}^{T}(\bar{s})\bar{f}(\bar{s})-\left(1+\alpha-\frac{\sigma}{2}\right)\bar{s}^{T}\bar{f}(\bar{s})\right)\\
    &+(-\gamma g_{a}\bar{s}^{T}\bar{f}(\bar{s})+(1-\varepsilon)(1+\alpha)\|W\|\|\bar{s}\|\|\bar{f}(\bar{s})\|\\
    &-(1-\varepsilon)(1+\alpha)\bar{s}^{T}\bar{s})+(-\delta g_{a}\bar{s}^{T}\bar{f}(\bar{s})\\
    &+(1-\varepsilon)\|W\|\|\bar{a}\|\|\bar{f}(\bar{s})\|-\varepsilon \bar{a}^{T}\bar{a}).
\end{aligned}
\end{equation}
By the fact that $2\bar{f}^{T}(\bar{s})\bar{f}(\bar{s})\leq \bar{s}^{T}\bar{f}(\bar{s})$, and $\sigma=2(1+\alpha)-\mu_{max}(W)$, we have from (\ref{eq36})
\begin{equation}
\begin{aligned}\label{eq37}
\dot{V}&\leq (-2\gamma g_{a}\bar{f}^{T}(\bar{s})\bar{f}(\bar{s})+(1-\varepsilon)(1+\alpha)\|W\|\|\bar{s}\|\|\bar{f}(\bar{s})\|\\
    &-(1-\varepsilon)(1+\alpha)\bar{s}^{T}\bar{s})+(-2\delta g_{a}\bar{f}^{T}(\bar{s})\bar{f}(\bar{s})\\
    &+(1-\varepsilon)\|W\|\|\bar{a}\|\|\bar{f}(\bar{s})\|-\varepsilon \bar{a}^{T}\bar{a}).
\end{aligned}
\end{equation}

By the equation (\ref{eq37}), it is clear that $\dot{V}$ is negative definite when $g_{a}$ satisfies
\begin{equation}
  \nonumber
  g_{a}>\max\left(\frac{(1-\varepsilon)(1+\alpha)\|W\|^2}{8\gamma},\frac{(1-\varepsilon)^{2}\|W\|^2}{8\varepsilon\delta}\right),
\end{equation}
which can be easily verified by the definition $g_{a}=\bar{g}_{a}/\alpha$ and the equations (\ref{eq30}) (\ref{eq34}) and (\ref{eq35}).

This completes the proof of this proposition.      $\blacksquare$

Theorem 3 indicates that if $\mu_{1}$ is slightly larger than $m(1+\alpha)$, then the system (\ref{eq7}) may have a limit cycle attractor, which might be a recalled pattern. Proposition 2 indicates that if $\mu_{max}(W)$ is less than $2(1+\alpha)$, and $\bar{g}_{a}$ is large enough, then unfortunately, the system (\ref{eq7}) may have a globally asymptotically stable equilibrium consisting of synchronized minicolumn states in each hypercolumn and there might not be a recalled pattern in the free recall model.

We summarize the results of Theorem 3 and Proposition 2 in the Table~\ref{tab1}.

\section{Simulation Results}

In this section, we will verify theoretical results in Section 3 by numerical simulations.

Set $n=6$, $m=3$, and $\alpha ,\bar{g}_{a}$ are taken as $1/54$ and $97/54$, respectively(see, \cite{Villani2020}). As discussed in Section 2, the connection weight matrix $W$ in the system (\ref{eq7}) is determined by the patterns encoded in the model previously and the learning rule. We assume that three patterns to be stored and recalled are $z^{(1)}=[1,1,1,1,1,1]$(pattern 1), $z^{(2)}=[2,2,2,1,1,1]$(pattern 2) and $z^{(3)}=[2,2,3,1,3,2]$(pattern 3). For simplicity, we use the following learning rule.

\begin{learning rule} 

  Step 1: For pattern $r$, set
\begin{equation}\label{eq65}
  \bar{w}^{(r)}_{ij,kl}=
    \begin{cases}
        1, & \mbox{if } j=z^{(r)}_{i}, l=z^{(r)}_{k}, \\
         -\frac{1}{m-2},  &\mbox{if } j=z^{(r)}_{i},l\neq z^{(r)}_{k},\\
         -\frac{1}{m-2},  &\mbox{if } j\neq z^{(r)}_{i}, l= z^{(r)}_{k}, \\
        0, & \mbox{otherwise},
    \end{cases}
\end{equation}
where $z^{(r)}_{i}$ is the $i_{th}$ element of pattern $z^{(r)}$.

  Step 2: Compute
\begin{equation}\label{eq60}
  \bar{w}_{ij,kl}=\sum_{r}w^{(r)}_{ij,kl}, \qquad 1\leq i,k\leq n, 1\leq j,l\leq m.
\end{equation}

  Step 3: Rewrite $\bar{w}_{ij,kl}$ as an $mn\times mn$ matrix $\bar{W}$ by the way of the equation (\ref{eq4}). The connection weight matrix $W$ is given by $W=\mu_{1}\bar{W}_{0}$, where $\mu_{1}>0$, $\bar{W}_{0}=\frac{\bar{W}}{m\mu_{max}(\bar{W}\Lambda)}$, $\mu_{max}(\bar{W}\Lambda)$ is the largest eigenvalue of matrix $\bar{W}\Lambda$, and $\Lambda$ is defined in (\ref{eq10}).
  
\end{learning rule}

The above learning rule is a specific form of Hebb's rule. By the equation (\ref{eq65}) in the above Learning rule, if two minicolumns are activated simultaneously (in a pattern), we give a positive weight $1$ to the connection between them; if one minicolumn is activated while another minicolumn is not activated, we give a negative connection weight $-\frac{1}{m-2}$; otherwise, we give a zero connection weight.

For the connection weight matrix $W$ obtained from the above learning rule, it is clear that Assumptions 1-3 are satisfied. By Lemma 2, the equation (\ref{eq95}), the fact $tr(\bar{W})=0$ and the fact $\sum_{l=1}^{m}\bar{w}^{(r)}_{ij,kl}=-\frac{1}{m-2}$ for each $i,j,k,r$, we have $\mu_{max}(\bar{W}\Lambda)>0$ and $\mu_{max}(W)=\mu_{1}$, where $tr(\bar{W})$ is the trace of the matrix $\bar{W}$ and $\mu_{max}(W)$ is the largest eigenvalue of $W$. A pattern is recalled means that the corresponding minicolumn in each hypercolumns is activated, namely, the output of the corresponding minicolumn in each hypercolumn is larger than a threshold which is close to 1. For example, pattern 3 $z^{(3)}=[2,2,3,1,3,2]$ is recalled means that $o_{1,2}$, $o_{2,2}$, $o_{3,3}$, $o_{4,1}$, $o_{5,3}$, and $o_{6,2}$ are all larger than the threshold. We take the threshold as 0.9.

As the connection weight matrix $W$ is determined, by simulation method, we can observe whether there are recalled patterns in the system (\ref{eq7}) when $\mu_{1}$ takes different values.

\begin{figure}[!t]
\centering
\includegraphics[width=2.5in]{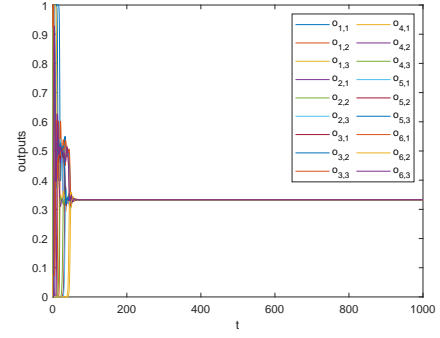}
\caption{The outputs of the system (\ref{eq7}) when $\mu_{1}=2(1+\alpha)-0.1$.}
\label{fig6}
\end{figure}

Fig.~\ref{fig6} shows dynamics of the outputs of the system (\ref{eq7}) when $\mu_{1}=2(1+\alpha)-0.1$.

In Fig.~\ref{fig6}, we see that all of the outputs $o_{i,j} (1\leq i\leq 6, 1\leq j\leq 3)$ converge to $\frac{1}{3}$, which indicates that no pattern is recalled.

Fig.~\ref{fig7} shows dynamics of the outputs of the system (\ref{eq7}) when $\mu_{1}=3(1+\alpha)+40$.


The attractor in Fig.~\ref{fig7} is a periodic orbit whose period is about $59$. In Fig.~\ref{fig7}, we see that three patterns might be recalled.

When $\mu_{1}=3(1+\alpha)+200$ so that the condition of Proposition 1 is not satisfied, the system (\ref{eq7}) might have more than one equilibrium. Fig.~\ref{fig8} shows dynamics of the outputs of the system (\ref{eq7}) with three different initial value when $\mu_{1}=3(1+\alpha)+200$.

In Fig.~\ref{fig8}, we see that with appropriate initial values, all of three patterns can be recalled.

When $\bar{g}_{a}$ is taken as $50$, $\mu_{1}=3(1+\alpha)+40$, the system converges to an attractor which is neither equilibrium nor limit cycle. We calculate Lyapunov exponents of the attractor in this case by a numerical method in \cite{Wolf1985}. All of Lyapunov exponents are displayed in Fig.~\ref{fig9}. It is proved in \cite{Eckmann1985} that an attractor different from equilibria must have a zero Lyapunov exponent. The third Lyapunov exponent of the attractor in this system is 0 ($\pm 1\times 10^{-4}$), in accord with the above assertion. The largest Lyapunov exponent is 0.163($\pm 1\times 10^{-3}$). The fact that the attractor has positive Lyapunov exponents indicates that this is a strange attractor(cf.,\cite{Eckmann1985}).


Fig.~\ref{fig10} displays this attractor by showing the outputs of minicolumns which are activated when patterns are recalled when the time $t\in [4950,4970]$.

In Fig.~\ref{fig10}, we see that three patterns might be recalled under this condition.

\section{Conclusion}

This paper discusses the steady state behavior of a free recall model which consists of $n$ hypercolumns where $m$ minicolumns are included. Under mild conditions on the connection weights, a sufficient condition under which the free recall model has a limit cycle attractor is obtained by the bifurcation method. Besides, we give a sufficient condition where the origin is a globally asymptotically stable equilibrium of the free recall model, which indicates that no stored pattern is recalled. Finally, we find that besides equilibria and limit cycles, patterns can also be stored in a strange attractor (or chaos). An example in which the free recall model converges to a strange attractor (or chaos) where patterns are recalled is given to support this statement.



\appendix
\section{Algorithm to compute $\frac{d^{3}\tilde{V}}{dv_{1}^{3}}(0)$}    
The following algorithm can be used to compute $\frac{d^{3}\tilde{V}}{dv_{1}^{3}}(0)$, where $\tilde{V}$ is defined in (\ref{eq14}).

\begin{algorithm}(cf., \cite{Marsden1976})
  Step 1:  Compute $d_{i}d_{j}g(0,0),i,j\in\{1,2\}$ by
  \begin{equation}
  \begin{aligned}\label{eq38}
   &\left[\begin{array}{ccc}
           d_{3}X^{3}_{0}(0,0,0) & 2|\lambda(0)|I & O \\
           -|\lambda(0)|I & d_{3}X^{3}_{0}(0,0,0) & |\lambda(0)|I \\
           O & -2|\lambda(0)|I &  d_{3}X^{3}_{0}(0,0,0)
         \end{array}
   \right] \\
   &\times \left[\begin{array}{c} d_{1}d_{1}g(0,0) \\ d_{1}d_{2}g(0,0)\\ d_{2}d_{2}g(0,0) \end{array}\right]=\left[\begin{array}{c} -d_{1}d_{1}X^{3}_{0}(0,0,0) \\ -d_{1}d_{2}X^{3}_{0}(0,0,0)\\ -d_{2}d_{2}X^{3}_{0}(0,0,0) \end{array}\right],
  \end{aligned}
  \end{equation}
  where $d_{i}=\frac{\partial}{\partial v_{i}}$.

  Step 2: Compute the second and third derivatives of $\hat{X}_{0}(0,0)$ by the following equations.
  \begin{equation}\label{eq39}
    d_{k}d_{j}\hat{X}^{i}_{0}(0,0)=d_{k}d_{j}X^{i}_{0}(0,0,0) \qquad  i,j,k = 1,2.
  \end{equation}
  \begin{equation}\label{eq40}
  \begin{aligned}
    d_{l}d_{k}d_{j}\hat{X}^{i}_{0}(0,0)&=d_{l}d_{k}d_{j}X^{i}_{0}(0,0,0)\\
    &+d_{3}d_{j}X^{i}_{0}(0,0,0)\circ d_{l}d_{k}g(0,0)\\
    &+d_{3}d_{k}X^{i}_{0}(0,0,0)\circ d_{l}d_{j}g(0,0)\\
    &+d_{3}d_{l}X^{i}_{0}(0,0,0)\circ d_{k}d_{j}g(0,0) \\
    &i,j,k,l = 1,2.
  \end{aligned}
  \end{equation}

  Step 3: Compute $\frac{d^{3}\tilde{V}}{dv_{1}^{3}}(0)$ by
  \begin{equation}
  \begin{aligned}\label{eq41}
    &\frac{d^{3}\tilde{V}}{dv_{1}^{3}}(0)
    =\frac{3\pi}{4|\lambda(0)|}
    (d^{3}_{1}\hat{X}^{1}_{0}(0,0)
    +d_{1}d^{2}_{2}\hat{X}^{1}_{0}(0,0)\\
    &+d^{2}_{1}d_{2}\hat{X}^{2}_{0}(0,0)
    +d^{3}_{2}\hat{X}^{2}_{0}(0,0))\\
    &+\frac{3\pi}{4|\lambda(0)|^{2}}
    (-d^{2}_{1}\hat{X}^{1}_{0}(0,0)\circ d_{1}d_{2}\hat{X}^{1}_{0}(0,0)\\
    &+d^{2}_{2}\hat{X}^{2}_{0}(0,0)\circ d_{1}d_{2}\hat{X}^{2}_{0}(0,0)
    +d^{2}_{1}\hat{X}^{2}_{0}(0,0)\circ d_{1}d_{2}\hat{X}^{2}_{0}(0,0)\\
    &-d^{2}_{2}\hat{X}^{1}_{0}(0,0)\circ d_{1}d_{2}\hat{X}^{1}_{0}(0,0)
    +d^{2}_{1}\hat{X}^{1}_{0}(0,0)\circ d^{2}_{1}\hat{X}^{2}_{0}(0,0)\\
    &-d^{2}_{2}\hat{X}^{1}_{0}(0,0)\circ d^{2}_{2}\hat{X}^{2}_{0}(0,0)).
  \end{aligned}
  \end{equation}
\end{algorithm}

\begin{figure}[!t]
	\centering
	\subfloat[]{ \includegraphics[width=2.5in]{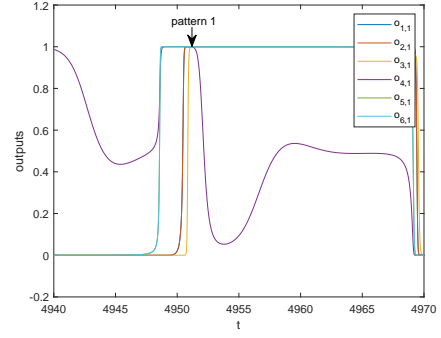} \label{fig7_1}}
	\quad
	\subfloat[]{ \includegraphics[width=2.5in]{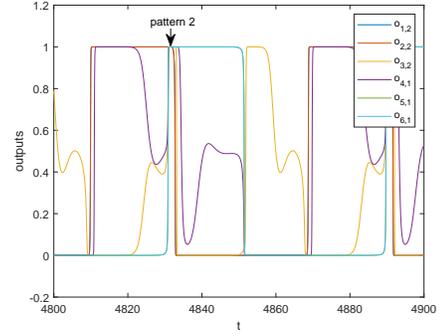} \label{fig7_2}}
	
	\quad
	\subfloat[]{ \includegraphics[width=2.5in]{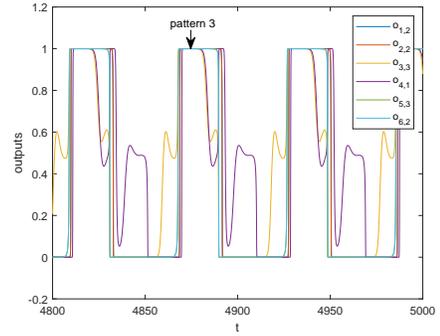} \label{fig7_3}}
	\caption{The outputs of the system (\ref{eq7}) when $\mu_{1}=3(1+\alpha)+40$. (a) Outputs of minicolumns which are activated when pattern 1 is recalled. If all the outputs in this figure are larger than the threshold 0.9, then pattern 1 is recalled. (b) Outputs of minicolumns which are activated when pattern 2 is recalled. If all the outputs in this figure are larger than the threshold 0.9, then pattern 2 is recalled. (c) Outputs of minicolumns which are activated when pattern 3 is recalled. If all the outputs in this figure are larger than the threshold 0.9, then pattern 3 is recalled.}
	\label{fig7}
\end{figure}

\section{Proof of $\frac{d^{3}\tilde{V}}{dv_{1}^{3}}(0)<0$}         
As $\tilde{V}$ is defined when the bifurcation parameter $\kappa=0$, in the following computation of $\frac{d^{3}\tilde{V}}{dv_{1}^{3}}(0)$, we have $\mu_{1}=m(1+\alpha)$ and $\mu_{i}<\mu_{1}=m(1+\alpha)$ for $2\leq i\leq (m-1)n$, where $\mu_{i}$ are defined in Lemma 2. It is clear that we can define a $2mn\times 2mn$ invertible matrix $Q$ of the following form,

\begin{figure}[!t]
	\centering
	\subfloat[]{ \includegraphics[width=2.5in]{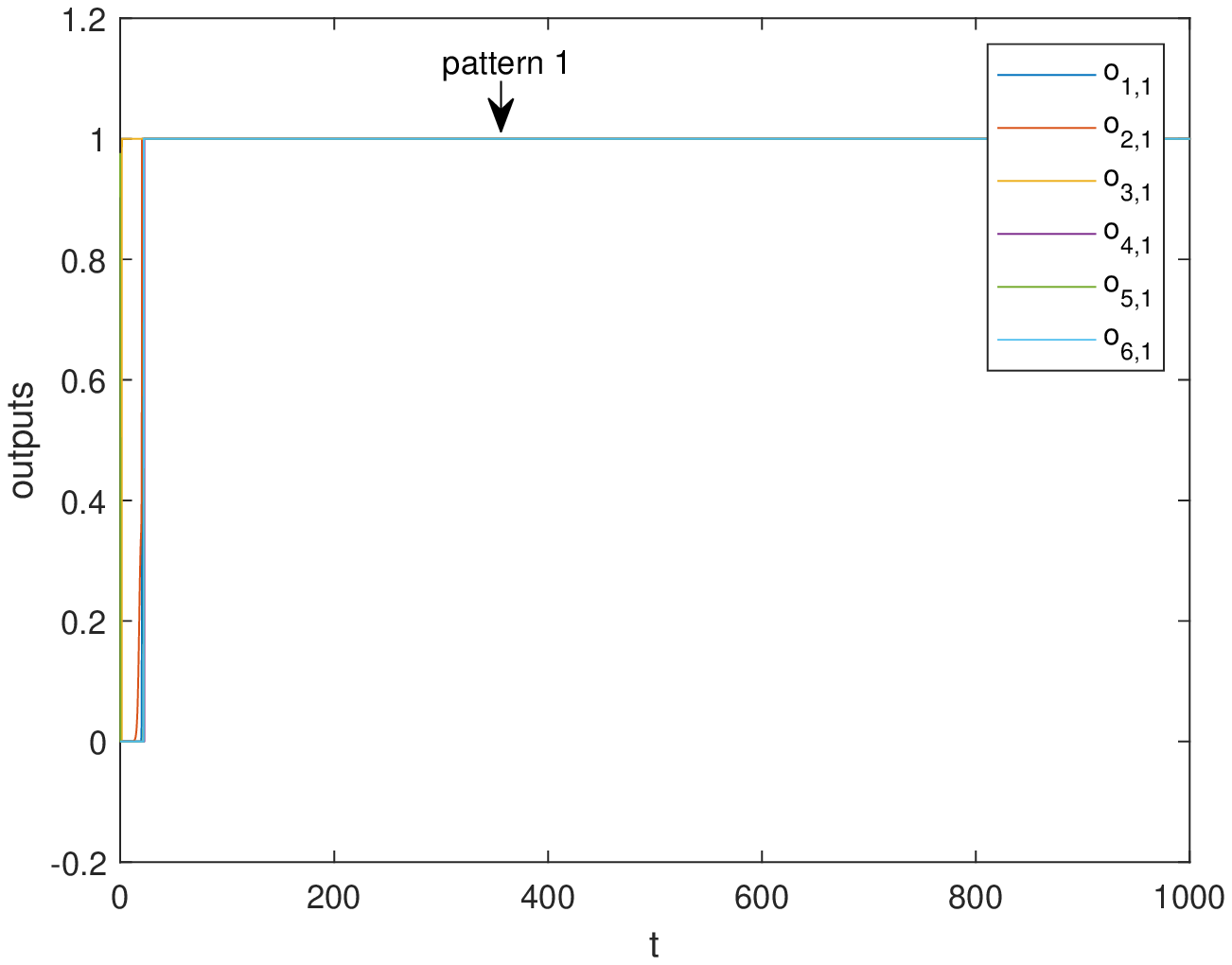} \label{fig8_1}}
	\quad
	\subfloat[]{ \includegraphics[width=2.5in]{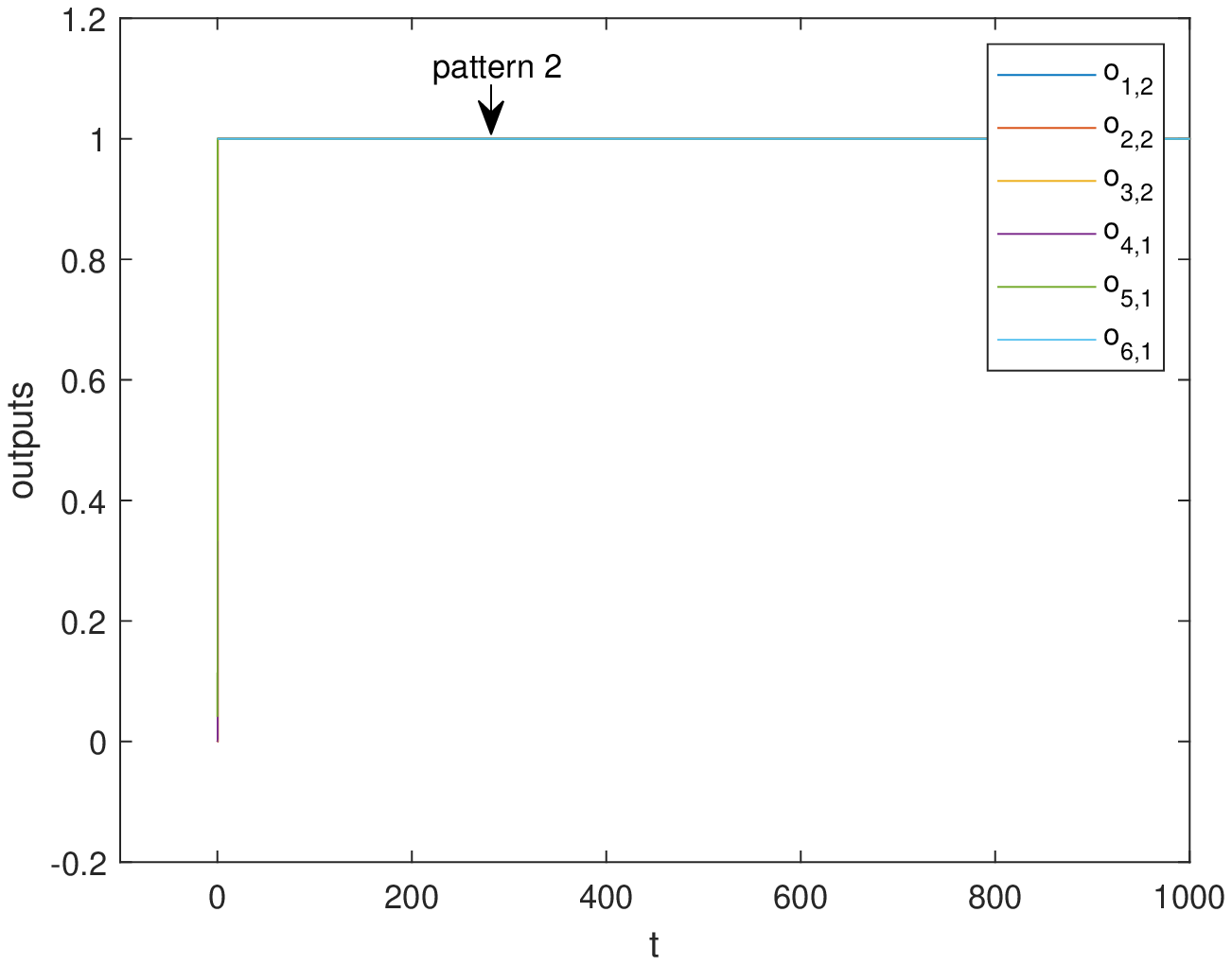} \label{fig8_2}}
	
	\quad
	\subfloat[]{ \includegraphics[width=2.5in]{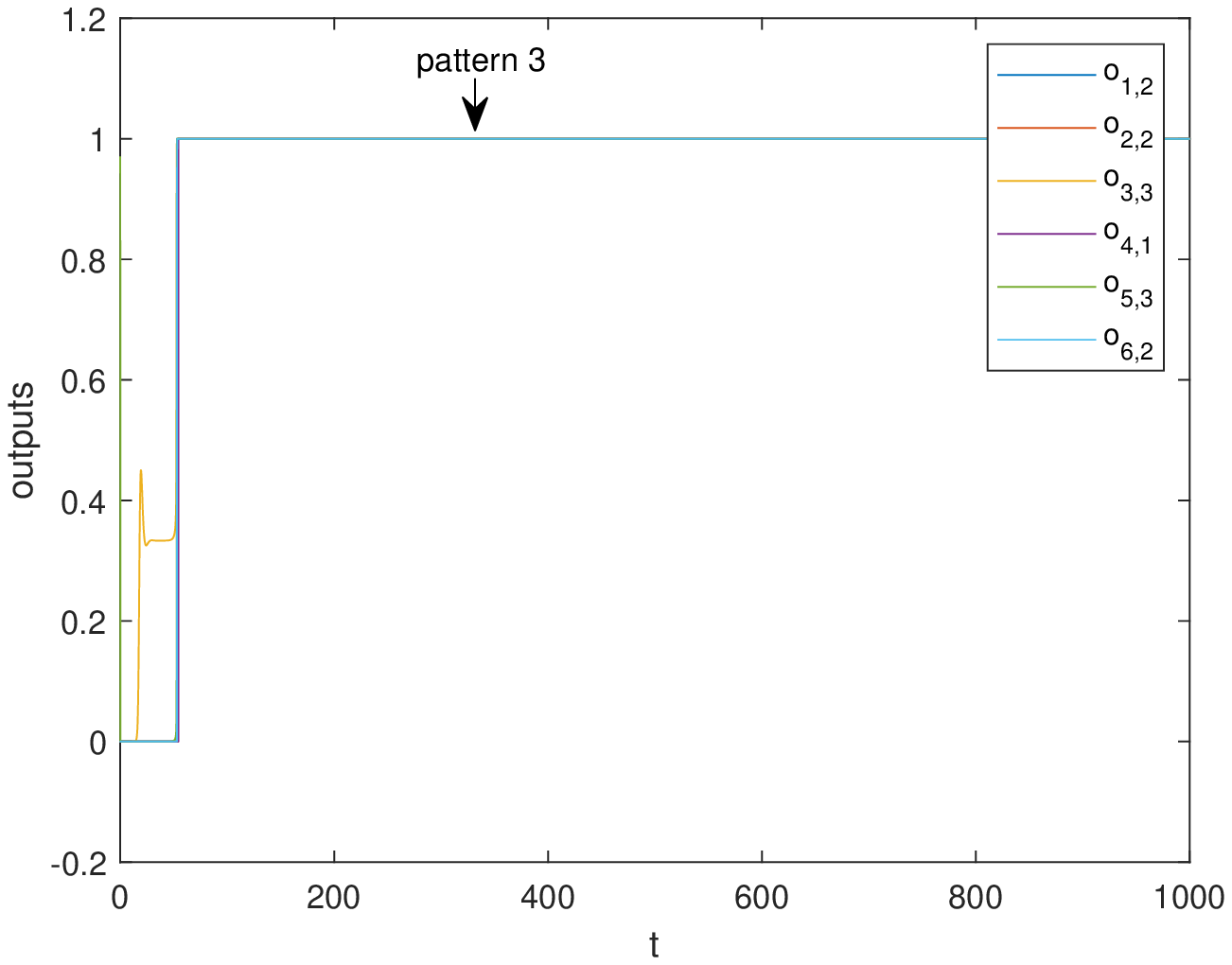} \label{fig8_3}}
	\caption{The outputs of the system (\ref{eq7}) with three different initial value when $\mu_{1}=3(1+\alpha)+200$. (a) Outputs of minicolumns which are activated when pattern 1 is recalled. If all the outputs in this figure are larger than the threshold 0.9, then pattern 1 is recalled. (b) Outputs of minicolumns which are activated when pattern 2 is recalled. If all the outputs in this figure are larger than the threshold 0.9, then pattern 2 is recalled. (c) Outputs of minicolumns which are activated when pattern 3 is recalled. If all the outputs in this figure are larger than the threshold 0.9, then pattern 3 is recalled.}
	\label{fig8}
\end{figure}

\begin{equation}
  \nonumber
  Q = \begin{bmatrix}
        Q_{1} &   &   &   \\
          & Q_{2} &   &   \\
          &   & \ddots &   \\
          &   &   & Q_{mn}
      \end{bmatrix},
\end{equation}
where $Q_{i} (1\leq i\leq mn)$ is a $2\times2$ invertible matrix which satisfies
\begin{equation}
  \nonumber
  Q_{1}^{-1}\left[\begin{array}{cc}\alpha & -1 \\ \frac{\bar{g}_a}{m} & -\alpha\end{array}\right] Q_{1} =\left[\begin{array}{cc} 0 & |\lambda(0)| \\ -|\lambda(0)| & 0\end{array}\right],
\end{equation}
\begin{equation}
  \label{eq79}
  \begin{aligned}
    & Q_{i}^{-1}\left[\begin{array}{cc}-1+\frac{\mu_{i}}{m} & -1 \\ \frac{\bar{g}_a}{m} & -\alpha\end{array}\right] Q_{i} \\
    & =\left[\begin{array}{cc} \nu_{i,+} & 0\\ 0 & \nu_{i,-}\end{array}\right], \qquad 2\leq i \leq (m-1)n,
  \end{aligned}
\end{equation}
\begin{equation}
  \nonumber
  \begin{aligned}
  & Q_{i}^{-1}\left[\begin{array}{cc}-1 & -1 \\ 0 & -\alpha\end{array}\right] Q_{i} \\
  & =\left[\begin{array}{cc} \nu_{i,+} & 0\\ 0 & \nu_{i,-} \end{array}\right], \qquad  (m-1)n+1\leq i \leq mn,
  \end{aligned}
\end{equation}
$\nu_{i,+}$ and $\nu_{i,-}$ are defined in (\ref{eq90}) and (\ref{eq12}).

In order to calculate $\frac{d^{3}\tilde{V}}{dv_{1}^{3}}(0)$, we introduce the following linear transformation of the system (\ref{eq7}) so that in this coordinate system, the equation (\ref{eq13}) can be satisfied,
\begin{equation}
  \nonumber
  \left[\begin{array}{c}\bar{s} \\ \bar{a} \end{array}\right]=\bar{P}TQv,
\end{equation}
where $v$ is a $2mn$ dimension vector, $Q$ is defined above, $T=[\varepsilon_{1},\varepsilon_{mn+1},\varepsilon_{2},\varepsilon_{mn+2},\cdots,\varepsilon_{mn},\varepsilon_{2mn}]$ is a $2mn\times 2mn$ matrix, $\varepsilon_{i}$ is the $i_{th}$ column of $2mn\times 2mn$ unit matrix, and
\begin{equation}
  \nonumber
  \bar{P} = \left[\begin{array}{cc}P & O \\ O & P\end{array}\right],
\end{equation}
where $P$ is defined in Lemma 2.

By this transformation and Lemma 2, for $\kappa=0$, that is $\mu_{1}=m(1+\alpha)$, we get the following system from (\ref{eq7})

\begin{equation}
\begin{aligned}\label{eq15}
  \dot{v}
  & = \bar{H}v+Q^{-1}T^{T}\left[\begin{array}{c}D_{3} \\ \bar{g}_{a}I \end{array} \right]\\
  & \times P^{T}(\bar{f}(\left[\begin{array}{cc}P&O\end{array}\right]TQv)-\Lambda \left[\begin{array}{cc}P&O\end{array}\right]TQv)\\
  & =X_{0}(v),
\end{aligned}
\end{equation}
where
\begin{equation}\nonumber
  \bar{H} = \begin{bmatrix}
              0 & \mid\lambda(0)\mid & O  \\
              -\mid\lambda(0)\mid & 0 & O   \\
              O  & O  & \bar{H}_{3}
            \end{bmatrix},
\end{equation}
\begin{equation}\label{eq16}
\bar{H}_{3}=\begin{bmatrix}
                  \nu_{2,+} &   &   &   &   \\
                    & \nu_{2,-} &   &   &   \\
                    &   & \ddots &   &   \\
                    &   &   & \nu_{mn,+} &   \\
                    &   &   &   & \nu_{mn,+}
                \end{bmatrix},
\end{equation}
and $D_{3}$ is defined in Lemma 2.

\begin{figure}[!t]
\centering
\includegraphics[width=2.5in]{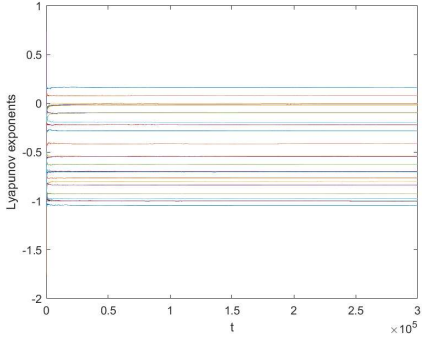}
\caption{The Lyapunov exponents of the attractor of the system (\ref{eq7}) when $\bar{g}_{a}=50, \mu_{1}=3(1+\alpha)+40$.}
\label{fig9}
\end{figure}

We next calculate $\frac{d^{3}\tilde{V}}{dv_{1}^{3}}(0)$ by Algorithm 1 in Appendix.

In step 1, we denote $X_{0}(v)=[X^{1}_{0}(v),X^{2}_{0}(v),(X^{3}_{0}(v))^{T}]^{T}$, $v=[v_{1},v_{2},v^{T}_{3}]^{T}$, $v_{3}=[\bar{v}_{3},\bar{v}_{4},\cdots,\bar{v}_{2mn}]^{T}$ and the differential operator $d_{i}=\frac{\partial}{\partial v_{i}}$, where $X^{1}_{0}(v),X^{2}_{0}(v),v_{1},v_{2} \in \mathbb{R}$ and $X^{3}_{0}(v),v_{3} \in \mathbb{R}^{2(mn-1)}$. We have
\begin{equation}\label{eq17}
  d_{3}X^{3}_{0}(v)|_{v=0}= \bar{H}_{3}.
\end{equation}

By the equations (\ref{eq38}) and (\ref{eq17}),we have
\begin{equation}\label{eq77}
\begin{aligned}
    &g_{11}=\bar{H}_{3}^{-1}(-d_{1}d_{1}X^{3}_{0}|_{v=0}-2\mid\lambda(0)\mid g_{12}), \\
    &g_{22}=\bar{H}_{3}^{-1}(-d_{2}d_{2}X^{3}_{0}|_{v=0}+2\mid\lambda(0)\mid g_{12}), \\
    &g_{12}=(\bar{H}_{3}^{2}+4\mid\lambda(0)\mid ^{2}I)^{-1} \\
    &\times(\mid\lambda(0)\mid( -d_{1}d_{1}X^{3}_{0}|_{v=0}+d_{2}d_{2}X^{3}_{0}|_{v=0})\\
    &-\bar{H}_{3}d_{1}d_{2}X^{3}_{0}|_{v=0}),
\end{aligned}
\end{equation}
where $g_{ij}=d_{i}d_{j}g(0,0)$ for $1\leq i,j\leq 2$.

In Step 2, we denote the nonlinear part of $X_{0}(v)$ by $\bar{X}_{0}(v)=[\bar{X}^{1}_{0}(v),\bar{X}^{2}_{0}(v),...,\bar{X}^{2mn}_{0}(v)]^{T}$, that is
\begin{equation}\label{eq18}
  \bar{X}_{0}(v)=Q^{-1}T^{T}\left[\begin{array}{c}D_{3} \\ \bar{g}_{a}I \end{array} \right]P^{T}\bar{f}(\left[\begin{array}{cc}P&O\end{array}\right]TQv).
\end{equation}
We know that second and third order derivatives of $X_{0}(v)$ are equal to those of $\bar{X}_{0}(v)$.
Denote $P=[p_{1},p_{2},\cdots, p_{mn}]$, 
\begin{equation}\nonumber
    p_{i}=[p_{i,11},p_{i,12},\cdots,p_{i,1m},p_{i,21},\cdots,p_{i,nm}]^{T},
\end{equation}

\begin{equation}\nonumber
  Q_{i}=\begin{bmatrix}
          q_{i,1} & q_{i,2} \\
          q_{i,3} & q_{i,4}
        \end{bmatrix},
\end{equation}
and
\begin{equation}\nonumber
  Q^{-1}_{i}=\begin{bmatrix}
          q^{(-1)}_{i,1} & q^{(-1)}_{i,2} \\
          q^{(-1)}_{i,3} & q^{(-1)}_{i,4}
        \end{bmatrix}.
\end{equation}

Denote $e_{i,1}=\mu_{i}q^{(-1)}_{i,1}+\bar{g}_{a}q^{(-1)}_{i,2}, e_{i,2}=\mu_{i}q^{(-1)}_{i,3}+\bar{g}_{a}q^{(-1)}_{i,4},1\leq i\leq mn$, where $\mu_{1}=m(1+\alpha)$ and $\mu_{i}<\mu_{1}=m(1+\alpha)$ for $2\leq i\leq (m-1)n$.

\begin{figure}[!t]
	\centering
	\subfloat[]{ \includegraphics[width=2.5in]{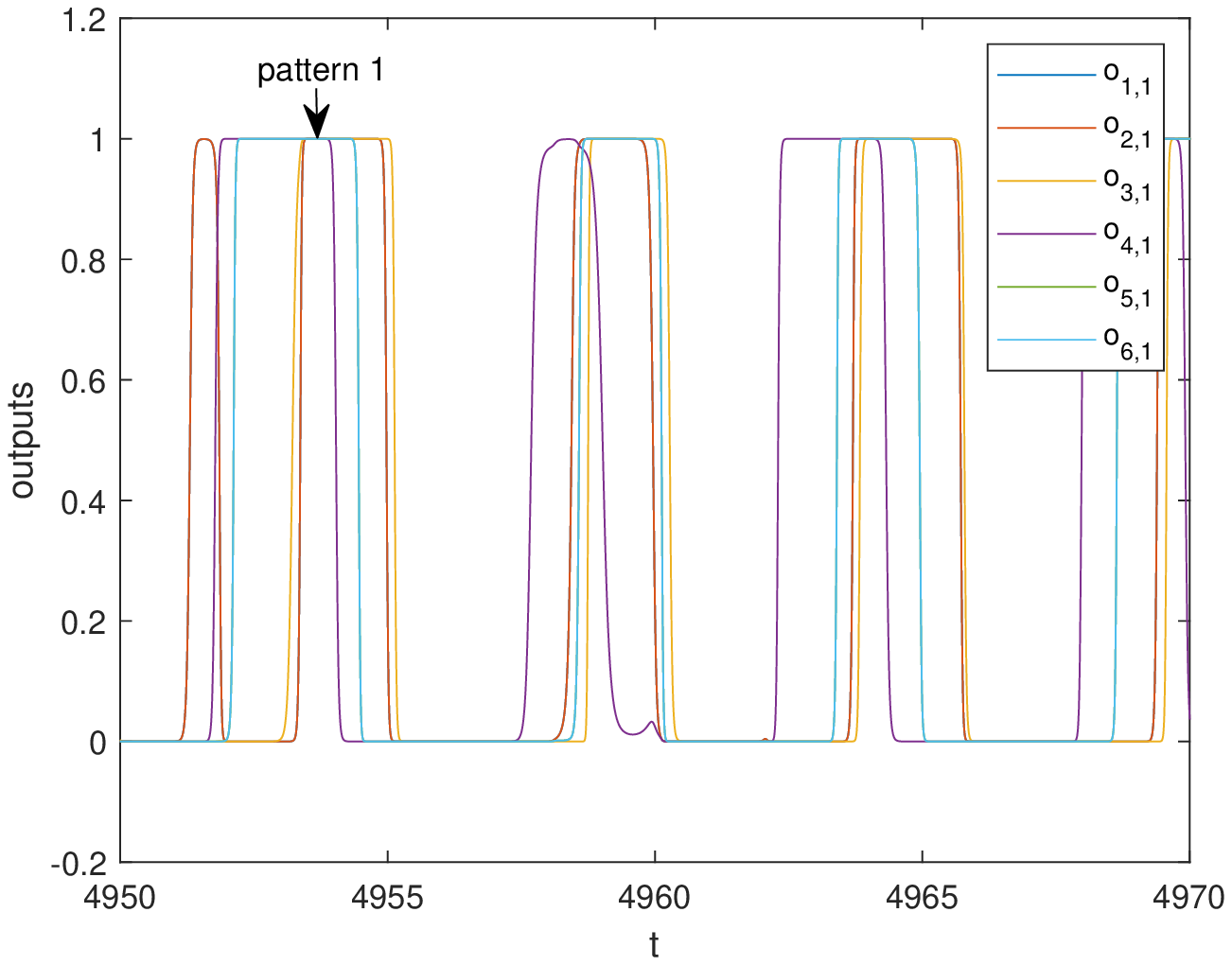} \label{fig10_1}}
	\quad
	\subfloat[]{ \includegraphics[width=2.5in]{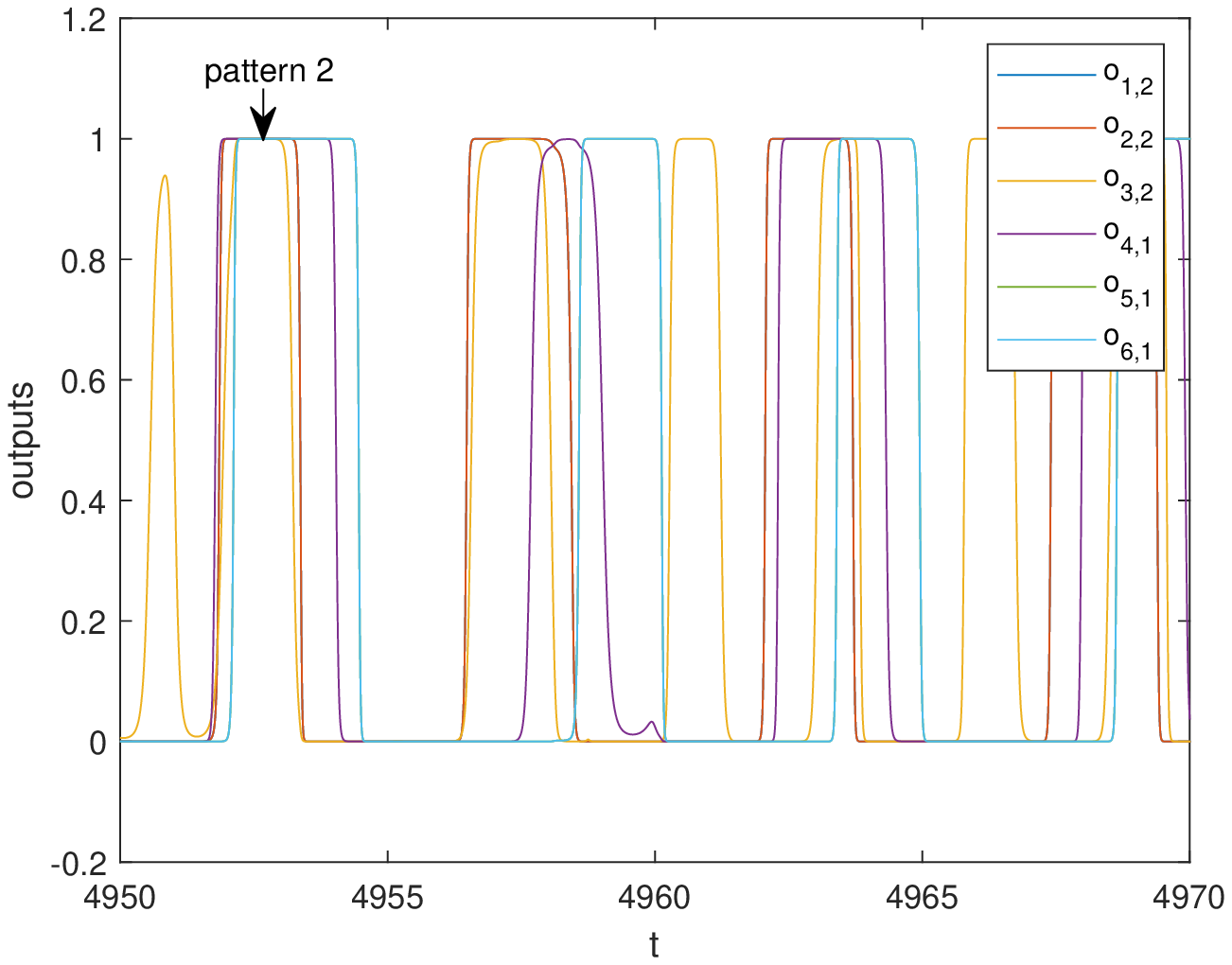} \label{fig10_2}}
	
	\quad
	\subfloat[]{ \includegraphics[width=2.5in]{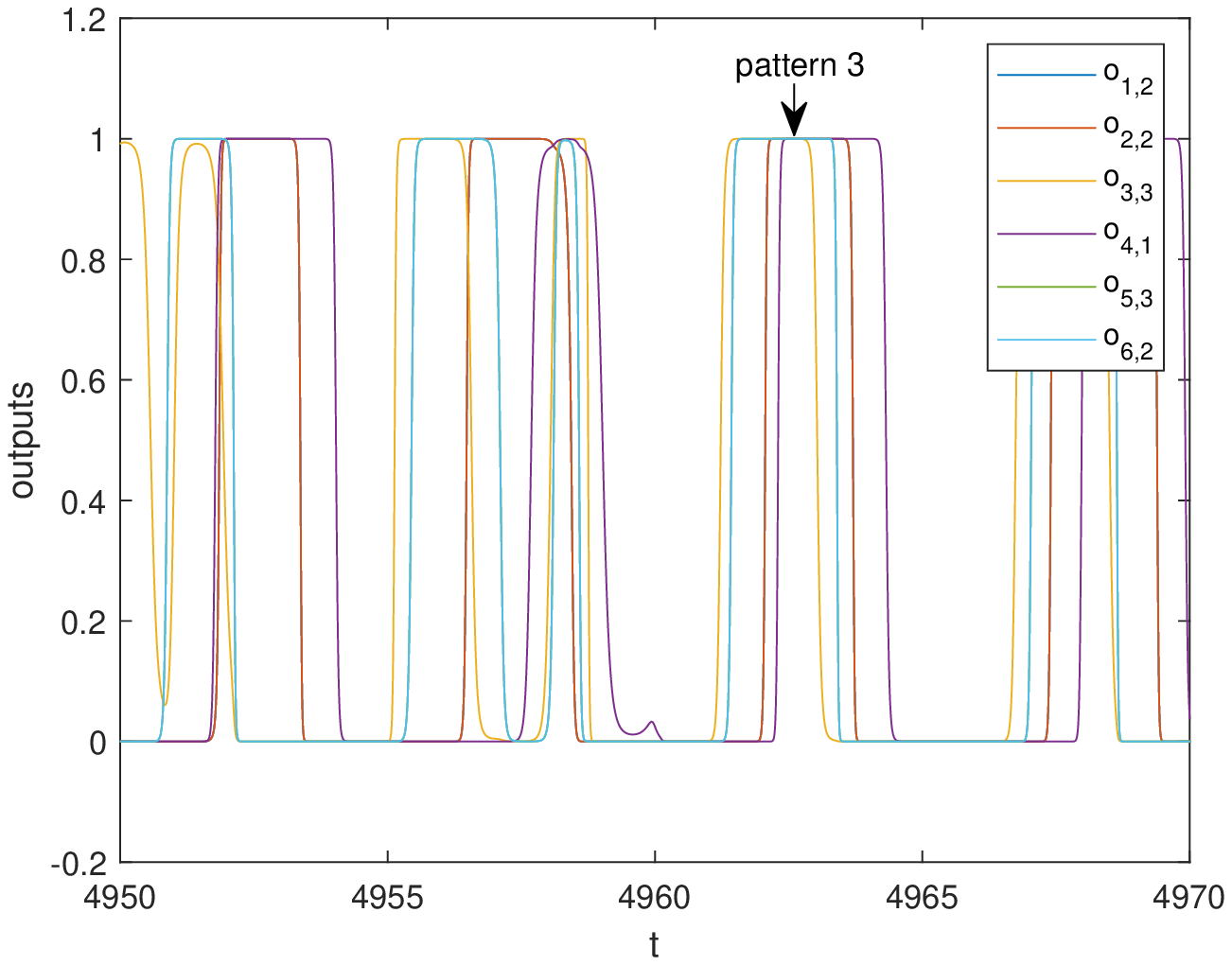} \label{fig10_3}}
	\caption{The outputs of the system (\ref{eq7}) with three different initial value when $\bar{g}_{a}=50, \mu_{1}=3(1+\alpha)+40$. (a) Outputs of minicolumns which are activated when pattern 1 is recalled. If all the outputs in this figure are larger than the threshold 0.9, then pattern 1 is recalled. (b) Outputs of minicolumns which are activated when pattern 2 is recalled. If all the outputs in this figure are larger than the threshold 0.9, then pattern 2 is recalled. (c) Outputs of minicolumns which are activated when pattern 3 is recalled. If all the outputs in this figure are larger than the threshold 0.9, then pattern 3 is recalled.}
	\label{fig10}
\end{figure}

By the equation (\ref{eq18}) and calculation, we have for $1\leq i\leq mn$
\begin{equation}\label{eq19}
\begin{aligned}
  &\bar{X}^{2i-1}_{0}(v)=e_{i,1}p^{T}_{i}\bar{f}(\sum_{k=1}^{mn}p_{k}(q_{k,1}v_{2k-1}+q_{k,2}v_{2k})), \\
  &\bar{X}^{2i}_{0}(v)=e_{i,2}p^{T}_{i}\bar{f}(\sum_{k=1}^{mn}p_{k}(q_{k,1}v_{2k-1}+q_{k,2}v_{2k})).
\end{aligned}
\end{equation}
We denote the $((r-1)m+t)_{th}$ element of $f(s)$ defined in (\ref{eq5}) as $f_{rt}(s)(1\leq r\leq n, 1\leq t\leq m)$. By the definition of $f(s)$, we have for $1\leq l\leq m$
\begin{equation}\label{eq70}
  \frac{\partial f_{rt}(s)}{\partial s_{rl}}= -f_{rt}(f_{rl}-\delta_{tl}),
\end{equation}
where $\delta_{tl}=1$ for $t=l$ and $\delta_{tl}=0$ for $t\neq l$.
By the equations (\ref{eq19}) and (\ref{eq70}), we have
\begin{equation}\label{eq71}
\begin{aligned}
  d_{1}^{2}\bar{X}^{1}_{0} & =e_{1,1}q_{1,1}^{2}\sum_{r,t}\sum_{k,l}p_{1,rt}p_{1,rk}p_{1,rl} \\
   & \times f_{rt}((f_{rl}-\delta_{tl})(f_{rk}-\delta_{tk})+f_{rk}(f_{rl}-\delta_{kl})).
\end{aligned}
\end{equation}
By the equation (\ref{eq71}), Remark 3, the fact $f_{rt}(0)=\frac{1}{m}$ and calculation, we have
\begin{equation}\nonumber
  d_{1}^{2}\bar{X}^{1}_{0}|_{v=0}=\frac{e_{1,1}q_{1,1}^{2}}{m}\sum_{r,t}p_{1,rt}^{3}.
\end{equation}
Similarly, we have
\begin{equation}\label{eq20}
  \begin{aligned}
    &d_{k}d_{j}\bar{X}^{2i-1}_{0}|_{v=0}=e_{i,1}q_{1,k}q_{1,j}u_{i},\\
    &d_{k}d_{j}\bar{X}^{2i}_{0}|_{v=0}=e_{i,2}q_{1,k}q_{1,j}u_{i},\\
    &for \qquad 1\leq j,k\leq 2, 1\leq i\leq mn.
  \end{aligned}
\end{equation}

\begin{equation}
  \begin{aligned}\label{eq21}
    &\bar{d}_{2i-1}d_{j}\bar{X}^{k}_{0}|_{v=0}=e_{1,k}q_{i,1}q_{1,j}u_{i},\\
    &\bar{d}_{2i}d_{j}\bar{X}^{k}_{0}|_{v=0}=e_{1,k}q_{i,2}q_{1,j}u_{i},\\
    &for \qquad 1\leq j,k\leq 2, 2\leq i\leq mn.
  \end{aligned}
\end{equation}
where $\bar{d}_{r}=\frac{\partial}{\partial \bar{v}_{r}}$ and
\begin{equation}\label{eq22}
\begin{aligned}
 & u_{i}=\frac{1}{m}\sum_{r=1}^{n}\sum_{t=1}^{m}p^{2}_{1,rt}p_{i,rt},\qquad 1\leq i\leq (m-1)n, \\
 & u_{i}=0, \qquad (m-1)n+1\leq i\leq mn.
\end{aligned}
\end{equation}
Besides, we have
\begin{equation}\label{eq23}
\begin{aligned}
  d_{l}d_{k}d_{j}\bar{X}^{i}_{0}|_{v=0} & =-\frac{ e_{1,i}q_{1,j}q_{1,k}q_{1,l}}{m} \\
   & \times (\frac{3}{m}\sum_{r=1}^{n}(\sum_{t=1}^{m}p^{2}_{1,rt})^{2}-\sum_{r=1}^{n}\sum_{t=1}^{m}p^{4}_{1,rt}) \\
   &  for \qquad 1\leq i,j,k,l\leq 2.
\end{aligned}
\end{equation}

In Step 3, by the following equations (\ref{eq39})-(\ref{eq41}), we obtain
\begin{equation}
  \label{eq84}
  \frac{d^{3}\tilde{V}}{dv_{1}^{3}}(0)=I_{1}+I_{2}+I_{3},
\end{equation}

where
\begin{equation}\label{eq72}
\begin{aligned}
    I_{1}&=\frac{3\pi}{4\mid\lambda(0)\mid}(d_{1}^{3}\bar{X}^{1}_{0}|_{v=0}+d_{1}d_{2}^{2}\bar{X}^{1}_{0}|_{v=0}\\
         &+d_{1}^{2}d_{2}\bar{X}^{2}_{0}|_{v=0}+d_{2}^{3}\bar{X}^{2}_{0}|_{v=0}),
\end{aligned}
\end{equation}

\begin{equation}\label{eq73}
\begin{aligned}
  I_{2}&= \frac{3\pi}{4\mid\lambda(0)\mid}((d_{3}d_{1}\bar{X}^{1}_{0}|_{v=0}\circ g_{11}+d_{3}d_{1}\bar{X}^{1}_{0}|_{v=0}\circ g_{22} \\
       &+d_{3}d_{2}\bar{X}^{2}_{0}|_{v=0}\circ g_{11}+d_{3}d_{2}\bar{X}^{2}_{0}|_{v=0}\circ g_{22})\\
       &+2(d_{3}d_{1}\bar{X}^{1}_{0}|_{v=0}\circ g_{11}+d_{3}d_{2}\bar{X}^{2}_{0}|_{v=0}\circ g_{22}\\
       &+d_{3}d_{2}\bar{X}^{1}_{0}|_{v=0}\circ g_{12}+d_{3}d_{1}\bar{X}^{2}_{0}|_{v=0}\circ g_{12})),
\end{aligned}
\end{equation}

\begin{equation}\label{eq74}
\begin{aligned}
  &I_{3}=\frac{3\pi}{4|\lambda(0)|^{2}}
    (-d^{2}_{1}\bar{X}^{1}_{0}|_{v=0}\circ d_{1}d_{2}\bar{X}^{1}_{0}|_{v=0}\\
    &+d^{2}_{2}\bar{X}^{2}_{0}|_{v=0}\circ d_{1}d_{2}\bar{X}^{2}_{0}|_{v=0}
    +d^{2}_{1}\bar{X}^{2}_{0}|_{v=0}\circ d_{1}d_{2}\bar{X}^{2}_{0}|_{v=0}\\
    &-d^{2}_{2}\bar{X}^{1}_{0}|_{v=0}\circ d_{1}d_{2}\bar{X}^{1}_{0}|_{v=0}
    +d^{2}_{1}\bar{X}^{1}_{0}|_{v=0}\circ d^{2}_{1}\bar{X}^{2}_{0}|_{v=0}\\
    &-d^{2}_{2}\bar{X}^{1}_{0}|_{v=0}\circ d^{2}_{2}\bar{X}^{2}_{0}|_{v=0}).
\end{aligned}
\end{equation}

By the equations (\ref{eq20}), (\ref{eq22}), (\ref{eq23}), (\ref{eq72}), (\ref{eq74}), the fact $e_{1,1}q_{1,1}+e_{1,2}q_{1,2}=\mu_{1}$, and $e_{1,1}q_{1,2}-e_{1,2}q_{1,1}=\frac{\bar{g}_{a}-m\alpha(1+\alpha)}{|\lambda(0)|}$, we have

\begin{equation}\label{eq85}
\begin{aligned}
  I_{1}&= \frac{3\pi}{4\mid\lambda(0)\mid}(-\frac{\mu_{1}(q_{1,1}^2+q_{1,2}^2)}{m}) \\
       & \times (\frac{3}{m}\sum_{r=1}^{n}(\sum_{t=1}^{m}p^{2}_{1,rt})^{2}-\sum_{r=1}^{n}\sum_{t=1}^{m}p^{4}_{1,rt}),
\end{aligned}
\end{equation}

\begin{equation}\label{eq86}
    I_{3}=\frac{3\pi u_{1}^{2}}{4\mid\lambda(0)\mid^{2}}(-\mu_{1}(q_{1,1}^2+q_{1,2}^2)\frac{\bar{g}_{a}-m\alpha (1+\alpha)}{\mid\lambda(0)\mid}),
\end{equation}

We next calculate $I_{2}$. For simplicity, we consider the case of $(m-1)n=2$ so that by (\ref{eq20})-(\ref{eq22}), only the first two elements of $d_{3}d_{j}\bar{X}^{i}_{0}|_{v=0}$ and $d_{i}d_{j}X^{3}_{0}|_{v=0}$ are not zero. We denote them by $\bar{d}_{3,4}d_{j}\bar{X}^{i}_{0}|_{v=0}$ and $d_{i}d_{j}X^{3,4}_{0}|_{v=0}$.

By (\ref{eq20}) and (\ref{eq21}), we have
\begin{equation}\label{eq75}
  d_{i}d_{j}X^{3,4}_{0}|_{v=0}=\begin{bmatrix}
    d_{i}d_{j}\bar{X}^{3}_{0}|_{v=0} \\
    d_{i}d_{j}\bar{X}^{4}_{0}|_{v=0}
  \end{bmatrix}=q_{1,i}q_{1,j}u_{2}\vec{l},
\end{equation}
where $\vec{l}=[e_{2,1},e_{2,2}]^{T}$, and
\begin{equation}\label{eq76}
  \bar{d}_{3,4}d_{j}\bar{X}^{i}_{0}|_{v=0}=\begin{bmatrix}
    \bar{d}_{3}d_{j}\bar{X}^{i}_{0}|_{v=0} \\
    \bar{d}_{4}d_{j}\bar{X}^{i}_{0}|_{v=0}
  \end{bmatrix}=q_{1,j}e_{1,i}u_{2}\vec{t},
\end{equation}
where $\vec{t}=[q_{2,1},q_{2,2}]^{T}$.

By (\ref{eq77})(\ref{eq73})(\ref{eq75})(\ref{eq76}) and calculation, we have
\begin{equation}\label{eq78}
\begin{aligned}
  I_{2}&=(q_{1,1}^{2}+q_{1,2}^{2})u_{2}^{2}(-2\mu_{1}\vec{t}^{T}\bar{H}_{3}^{-1}\vec{l}\\
       &-\mu_{1}\vec{t}^{T}\bar{H}_{3}(\bar{H}_{3}^{2}+4|\lambda(0)|^{2})^{-1}\vec{l} \\
       &+2(\bar{g}_{a}-\alpha  \mu_{1})\vec{t}^{T}(\bar{H}_{3}^{2}+4|\lambda(0)|^{2})^{-1}\vec{l}),
\end{aligned}
\end{equation}
where $\bar{H}_{3}$ is defined in (\ref{eq16}).

By (\ref{eq79}), we have
\begin{equation}\label{eq80}
    \begin{aligned}
    & \begin{bmatrix}
          q_{2,1} & q_{2,2} \\
          q_{2,3} & q_{2,4}
        \end{bmatrix}\left[\begin{array}{cc} \nu_{2,+}^{-1} & 0\\ 0 & \nu_{2,-}^{-1}\end{array}\right] Q_{2}^{-1} \\
    & =\frac{1}{\nu_{2,+}\nu_{2,-}}\left[\begin{array}{cc} -\alpha & 1\\ -\frac{\bar{g}_{a}}{m} & \frac{\mu_{2}}{m}-1\end{array}\right].
  \end{aligned}
\end{equation}

By (\ref{eq80}) and the definition of $e_{i,j}$, we have
\begin{equation}\label{eq81}
\begin{aligned}
  \vec{t}^{T}\bar{H}_{3}\vec{l} &=\begin{bmatrix}
          q_{2,1} & q_{2,2}
        \end{bmatrix}\left[\begin{array}{cc} \nu_{2,+}^{-1} & 0\\ 0 & \nu_{2,-}^{-1}\end{array}\right] Q_{2}^{-1}
        \begin{bmatrix}
          \mu_{2}\\
          \bar{g}_{a}
        \end{bmatrix} \\
                                &=\frac{\bar{g}_{a}-\alpha \mu_{2}}{\nu_{2,+}\nu_{2,-}}.
\end{aligned}
\end{equation}

Similarly, we have
\begin{equation}\label{eq82}
\begin{aligned}
   & \vec{t}^{T}\bar{H}_{3}(\bar{H}_{3}^{2}+4|\lambda(0)|^{2})^{-1}\vec{l} \\
   & =\frac{\nu_{2,+}\nu_{2,-}(\bar{g}_{a}-\alpha \mu_{2})+4|\lambda(0)|^{2}(\mu_{2}(\frac{\mu_{2}}{m}-1)-\bar{g}_{a})}{(4\mid \lambda(0)\mid^{2}+\nu_{2,+}^{2})(4\mid \lambda(0)\mid^{2}+\nu_{2,-}^{2})},
\end{aligned}
\end{equation}

and
\begin{equation}\label{eq83}
\begin{aligned}
   & \vec{t}^{T}(\bar{H}_{3}^{2}+4|\lambda(0)|^{2})^{-1}\vec{l} \\
   & =\frac{\frac{\bar{g}_{a}}{m}(\mu_{2}-m(1+\alpha))+3|\lambda(0)|^{2}\mu_{2}}{(4\mid \lambda(0)\mid^{2}+\nu_{2,+}^{2})(4\mid \lambda(0)\mid^{2}+\nu_{2,-}^{2})}.
\end{aligned}
\end{equation}

By (\ref{eq78}), (\ref{eq81})-(\ref{eq83}) and calculation, we obtain $I_{2}$ in the case of $(m-1)n=2$.

By similar way, we obtain $I_{2}$ in general case,
\begin{equation}\label{eq87}
\begin{aligned}
  I_{2}&=\frac{3\pi}{4\mid\lambda(0)\mid}(q_{1,1}^2+q_{1,2}^2) \\
       & \times \sum_{i=2}^{(m-1)n}\frac{u_{i}^2}{\nu_{i,+}\nu_{i,-}(4\mid \lambda(0)\mid^{2}+\nu_{i,+}^{2})(4\mid \lambda(0)\mid^{2}+\nu_{i,-}^{2})}A_{i},
\end{aligned}
\end{equation}
where
\begin{equation}\nonumber
  A_{i}=a_{1}\bar{\mu}_{i}^{3}+a_{2}\bar{\mu}_{i}^{2}+a_{3}\bar{\mu}_{i}+a_{4},
\end{equation}
\begin{equation}\nonumber
  \bar{\mu}_{i}=m(1+\alpha)-\mu_{i}>0, \qquad 2\leq i\leq (m-1)n,
\end{equation}
\begin{equation}\nonumber
  \begin{aligned}
    a_{1}&=-\frac{3}{m^{2}}\alpha (\alpha +1)(4\bar{g}_{a}-3m\alpha^{2}),\\
    a_{2}&=-\frac{1}{m^{2}}((20\alpha +12)\bar{g}_{a}^2-m\alpha(12+59\alpha+53\alpha^{2})\bar{g}_{a}\\
         &+3m^{2}\alpha^{3}(1+\alpha)(11\alpha+3)),\\
    a_{3}&=-\frac{1}{m^{2}}(\bar{g}_{a}-m\alpha^2)(8\bar{g}_{a}^2-m(4+21\alpha+23\alpha^{2})\bar{g}_{a}\\
         &+3m^{2}\alpha^{2}(1+\alpha)(5\alpha+8)),\\
    a_{4}&=-\frac{9}{m}(\alpha+1)(\bar{g}_{a}-m\alpha^{2})^{2}(\bar{g}_{a}-m\alpha(1+\alpha)).
  \end{aligned}
\end{equation}

By (\ref{eq90}), (\ref{eq12}) and the fact that $\mu_{i}<\mu_{1}=m(1+\alpha)$ for $2\leq i\leq (m-1)n$, it is clear that $\nu_{i,+}\nu_{i,-}>0$ and $(4\mid \lambda(0)\mid^{2}+\nu_{i,+}^{2})(4\mid \lambda(0)\mid^{2}+\nu_{i,-}^{2})>0$ for $2\leq i\leq (m-1)n$.

Now, by (\ref{eq84}), (\ref{eq85}), (\ref{eq86}) and (\ref{eq87}), we can decide the sign of $\frac{d^{3}\tilde{V}}{dv_{1}^{3}}(0)$ under condition (a) (b) or (c).

(\uppercase\expandafter{\romannumeral1}) If the condition (a) is satisfied, that is $m=2$, then we have $\sum_{t=1}^{m}p^{2}_{1,rt}p_{i,rt}=0$ by the fact that $\sum_{t=1}^{2}p_{i,rt}=0$, for $1\leq i\leq (m-1)n,1\leq r\leq n$ (see Remark 3). Thus, $u_{i}=0$ and we have $I_{2}=I_{3}=0$. By calculation, we have
\begin{equation}\nonumber
  I_{1}=-\frac{3\pi}{2\mid\lambda(0)\mid}\mu_{1}(q_{1,1}^2+q_{1,2}^2)\sum_{r=1}^{n}p^{4}_{1,r1}<0.
\end{equation}
Therefore, $\frac{d^{3}\tilde{V}}{dv_{1}^{3}}(0)<0$.

(\uppercase\expandafter{\romannumeral2}) If the condition (b) is satisfied, then it is clear that $I_{1}\leq0, I_{3}\leq0$. We can verify that $a_{i}<0$ for $i=1,2,3,4$, so $I_{2}<0$. Thus $\frac{d^{3}\tilde{V}}{dv_{1}^{3}}(0)<0$.

(\uppercase\expandafter{\romannumeral3}) If the condition (c) is satisfied, it is clear that $\frac{d^{3}\tilde{V}}{dv_{1}^{3}}(0)<0$. $\blacksquare$

\bibliographystyle{plain}        
\bibliography{autosam}           

\end{document}